\documentclass[twocolumn]{aastex63}

\usepackage{xfrac}
\usepackage{float}
\usepackage{physics}
\usepackage{amssymb}
\usepackage{savesym}
\savesymbol{tablenum}
\usepackage{rotating}
\usepackage[section]{placeins}
\usepackage[noabbrev]{cleveref}
\usepackage{subfigure}
\usepackage{gensymb}
\usepackage{siunitx}
\usepackage{url}

\restoresymbol{SIX}{tablenum}
\newcommand{\tus}[1]{$_{\text{2}}$}

\newcommand{\change}[1]{\textcolor{black}{#1}}
\newcommand{\move}[1]{\textcolor{black}{#1}}
\newcommand{\new}[1]{\textcolor{black}{#1}} 

\shorttitle{MEGASIM: Detecting ETAs}
\shortauthors{Yeager, Golovich, \& Pruett}

\graphicspath{{./}{figures/}}
\begin{document}
\title{MEGASIM: Distribution and Detection of Earth Trojan Asteroids}

\correspondingauthor{Travis Yeager}
\email{yeagerastro@gmail.com}

\author[0000-0002-2582-0190]{Travis Yeager}
\affiliation{Lawrence Livermore National Laboratory \\
7000 East Ave, Livermore \\
Livermore, CA 94550, USA}

\author[0000-0003-2632-572X]{Nathan Golovich}
\affiliation{Lawrence Livermore National Laboratory \\
7000 East Ave, Livermore \\
Livermore, CA 94550, USA}

\author[0000-0002-2911-8657]{Kerianne Pruett}
\affiliation{Lawrence Livermore National Laboratory \\
7000 East Ave, Livermore \\
Livermore, CA 94550, USA}

\begin{abstract}
Using \change{N-body simulation results from} the MEGASIM \change{dataset}, we present spatial distributions of Earth Trojan Asteroids and assess the detectability of the population in current and next-generation ground-based astronomical surveys \citep{Yeager_2022_lifetimes}. Our high-fidelity Earth Trojan Asteroid (ETA) distribution maps show never-before-seen high-resolution spatial features that evolve over timescales up to 1 Gyr. The simulation was synchronized to \change{start times and timelines of} two \change{observational astronomy} surveys, 1) the Vera C. Rubin Observatory's Legacy Survey of Space and Time (LSST) and 2) the Zwicky Transient Facility (ZTF). \change{We calculate u}pper limits for the \change{number of ETAs} \change{potentially observable with both the ZTF and LSST surveys}. Due to the Yarkovsky Effect, \change{we find no stable ETAs on billion year timescales likely to be detected by any ETA survey, as no C-type or S-type ETAs (with $H<22$ and $H<24$, respectively) are likely to be} stable on billion year timescales\change{, and} ETAs large enough to remain stable on billion year timescales are very rare relative to the rest of the ETA population. \change{We find that a twilight ETA survey will not drastically increase the likelihood of individual ETA detection, but would provide orders of magnitude more observations of select ETA populations.} \change{The null detection to date from ZTF restricts the potential ETA population to hundreds of objects larger than 100 meters \citep[at $H\approx22$,][]{2021AJ....161..282L}, while a} null detection \change{by} LSST will \change{further} restrict th\change{e ETA} population to tens of objects larger than 100 meters\change{.}
\end{abstract}

\section{Introduction}
\indent Trojan asteroids\footnote{\change{The term} `Trojan' \change{generally refers to} objects co-orbiting \change{the Sun} with Jupiter; however, given the general nature of these minor \change{bodies} across the Solar System, we use the term `Trojan' generically \change{to} refer to which planet they co-orbit \change{the Sun with}.} are a class of co-orbiting objects that librate about the L4 or L5 Lagrangian points within a planet's orbit. Mapping the distributions of possible Earth Trojan Asteroids (ETAs) is important to guide observational searches\change{, as ETAs may answer pressing science questions about how our Solar System and Moon were formed}. Librations of ETAs \change{can} carry \change{an} asteroid nearly the entire distance \change{between} the planet \change{they are co-orbiting with, and its} L3 \change{Lagrange} point on the far side of the orbit behind the Sun. In \change{a} rotating \change{reference} frame, the gravitational potential lines look like tadpoles\change{, where these} tadpole orbits occupy a half-toroid\change{al} region, \change{which} includes either the L4 or L5 point \change{within}. The region of the sky where ETAs potentially exist is difficult to observe due to the low Solar elongation. The geometry also tends to have small reflection angles, making observations even more difficult from Earth. However, synoptic surveys such as those carried out by the Zwicky Transient Facility \citep[ZTF,][]{bellm2018zwicky} and the Vera C. Rubin Observatory \citep{2009arXiv0912.0201L, jones2020scientific} could still overcome the difficult observing conditions \change{using} numerous observations at the periphery of the ETA regime. Twilight surveys, in particular, could be powerful probes into the population \change{of ETAs of interest;} ZTF has been carrying out a twilight survey for years \citep[see e.g.,][]{2020AJ....159...70Y}, and there is \change{current} discussion \change{on} whether or not \change{the} Rubin Observatory's Legacy Survey of Space and Time (LSST) will or will not have a dedicated twilight survey \citep{2018arXiv181200466S}. 

\subsection{Scientific \change{Impact} of ETAs}
\indent There are several potential sourcing mechanisms for Trojan asteroids\change{: 1)} they may be formed in the early proto-planetary disk near a planet's L4 or L5 Lagrange point \citep{2020A&A...642A.224M}\change{, 2)} \change{they} may be captured debris produced from major planetary impacts\change{, and 3)} they may be sourced from perturbed populations of asteroids elsewhere in the Solar System \change{(}either during planetary migration or from ejected asteroids from the main-asteroid belt\change{)}. The first two mechanisms would have occurred early in the formation of the Solar System, and we consider these to be ``primordial'' in nature. For ETAs, primordial sourcing through the impact hypothesis may also explain lunar formation. The giant impact hypothesis has been proposed, \change{suggesting that a} Mars-sized proto-planet could have formed in a stable orbit among debris\change{,} at either Earth's L4 or L5 Lagrange point\change{,} and collided with proto-Earth \move{from a 1 AU orbit in the solar nebula} to form the \change{M}oon \citep{2005AJ....129.1724B}. \change{It is hypothesized that a} debris cloud from such a collision \change{could} have provided a primordial source of ETAs \citep{CANUP2004433}. If found, such a population of asteroids would offer a unique lens into the evolution of the early Solar System and the Earth\change{-}Moon system. The existence of primordial ETAs would offer \change{ideal} low\change{-}$\Delta v$ targets for rendezvous missions to study the early inner \change{S}olar \change{S}ystem\change{,} especially the Earth-Moon system \citep{2019NatAs...3..193M}. NASA\change{'s} Lucy mission will soon begin its 12-year journey to several families of Jupiter Trojans, which will \change{begin} to answer \change{many} of the scientific questions we have discussed \citep{2020A&A...633A.153Z}. While the Lucy mission is focused on primordial Jupiter Trojans, it \change{may} shed light on how scientifically rich these objects are, \change{validating} that future Earth Trojan missions will be important in understanding our Solar System.

\subsection{Observational \change{ETA S}earches}\label{sec:obs_search}
\indent Observers have searched for ETAs with a few relatively small astronomical twilight surveys \citep{1998Icar..136..154W, 2000DPS....32.1407C, 2020MNRAS.492.6105M, 2021AJ....161..282L}. \citet{2020MNRAS.492.6105M} \change{and} \citet{2021AJ....161..282L} were able to use a null detection to place competitive upper limits on the ETA population, predicting an L4 ETA population\change{(${N}_{ET}$)} of ${N}_{ET} < 1$ for \change{absolute magnitude} $H=14$, ${N}_{ET} < 10$ for $H=16$, and ${N}_{ET} < 938$ for $H=22$. Efforts have been proposed on how to best observe the ETA population utilizing Solar System exploration missions \citep{2018LPI....49.1149C, 2018LPI....49.1771Y, 2019NatAs...3..193M}, ground-based telescopes \citep{2012MNRAS.420L..28T, 2014MNRAS.437.4019T, 2018arXiv181200466S, 2020AJ....159...70Y}, and \change{a near-Earth object (NEO)} Surveyor \citet{2021DPS....5330616M} which is planned \change{to} launch this decade to search for potentially hazardous asteroids\change{,} including ETAs. 

\indent \move{Currently, there are two known \change{ETAs}: 2010 $TK_7$ and 2020 $XL_5$ \citep{2011Natur.475..481C,2021RNAAS...5...29D}\footnote{\change{Both known ETAs are at Earth's L4 Lagrange point. None have been found at L5.}}. \change{Both ETAs} librate far from the Lagrange points \change{(i.e., they} are transient in nature\change{),} and will only be co-orbiting with Earth for a short period of time compared \change{to} the age of the \change{S}olar \change{S}ystem. \change{Additionally,} both known ETAs are on tadpole orbits that librate about Earth's L4 Lagrange point, \change{traversing} much of the distance between Earth and Earth's L3 Lagrange point behind the Sun. The orbits of both ETAs are stable on the order of thousands of years, with 2020 $XL_5$ being the more stable of the two. 2020 $XL_5$ was shown to likely have been captured in the 15th century and \change{should} remain stable for \change{$\sim$4000} years \citep{2022NatCo..13..447S}.}

\subsection{Previous ETA \change{S}imulations}

\indent \move{We refer readers to our previous \change{work in} analyzing the MEGASIM \change{results (described in more detail in \S \ref{sec:model})} for \move{an in-depth analysis on} lifetime and stability of ETAs\change{,} including a detailed literature review \citep{Yeager_2022_lifetimes}}. Numerous \change{additional ETA detectability} studies have been carried out \change{using} numerical simulations\move{, to \change{understand and constrain ETA populations for guiding} observers} \citep[\change{e.g.,}][]{2000Icar..145...33W,10.1046/j.1365-8711.2000.03761.x,MORAIS20021,2012MNRAS.420L..28T,Zhuo2019A&A...622A..97Z,2021AJ....161..282L,2022PSJ.....3..121N}\change{, and} have typically centered around the questions of sourcing, lifetimes, stability zones, and observability of the stable population. \citet{2000Icar..145...33W} \move{produced spatial maps of time-averaged ETA positions} from simulations of two synthetic populations of Earth Trojans \change{by injecting} co-orbiting \move{Earth} asteroids uniformly within the region between \change{the} L2 and L3 points. When viewing \change{these} ETA orbits from Earth, the resulting spatial density peak is a few degrees displaced toward the Sun from the L4 and L5 Lagrange points.  \citet{10.1046/j.1365-8711.2000.03761.x} and \citet{MORAIS20021} found stable zones with inclination between 10\degree{} and 45\degree{}\change{,} with a wide range of ecliptic longitudes encapsulating the L4 and L5 points\change{,} but spanning tens of degrees to either side of L4 or L5.  \citet{2012MNRAS.420L..28T} modeled albedo, size, and orbital distributions \change{of ETAs} in an effort to define an `optimal' search strategy for \change{ETA detection}, \change{and} based on their simulations, suggest a twilight observing campaign to search the wide swath of area that the ETAs in their simulation cover. The only known asteroid at the time \change{of these simulations}, $2010TK_{7}$, would \change{have} appeared in such a survey \citep{2012MNRAS.420L..28T}. While currently known ETAs exhibit orbital stability on the order of thousands of years, objects librating about the Earth's L4 or L5 Lagrange points can be stable on the order of millions or billions of years \citep{Yeager_2022_lifetimes}. An ETA's ability to remain stable depends on several factors, e.g., spin rate, shape, size, albedo, density, and surface conductivity\change{, of which} play a role in non-gravitational forces such as the Yarkovsky Effect. Smaller ETAs experience larger effects from Solar radiation, thus driving them from the co-orbital regime more quickly. Simulations by \citep{Zhuo2019A&A...622A..97Z} find that only ETAs larger than $\sim$100 meters remain stable for 1 Gyr. \change{\citet{2021AJ....161..282L} placed upper limit constraints on the number of ETAs in L4, as described in \S \ref{sec:obs_search}.}

\indent In this paper, we study the MEGASIM data to assess the observability of ETAs in ongoing and upcoming synoptic surveys. In \S\ref{sec:model}, we give a brief summary of the simulation details, and we describe our method of determining the flux of individual asteroids in the simulation. In \S\ref{sec:maps}, we present spatial distributions and discuss translating our simulation output into observational surveys. In \S\ref{sec:results}, we present \change{our} results regarding the observability of ETAs in two synoptic surveys. Finally, in \S\ref{sec:conclusions}, we offer conclusions of our analysis.

\section{MEGASIM}\label{sec:model}
\indent The Multitudinous Earth Greek Asteroid Simulation (MEGASIM), announced first in \citet{Yeager_2022_rnaas}, \change{resulted in} 11.2 million ETA orbits initialized on trajectories that span beyond the expected ETA stability regime around the L4 Lagrange point. The orbits have now been carried out to a maximum of 4.5 billion years, allowing a study of the longevity of ETA orbits.

\subsection{Initialization and Parallelization}
\indent The MEGASIM data set consists of two distinct simulations\change{, one using the WHFast integrator \citep{wh, reboundwhfast}, and the other using the IAS15 \citep{reboundias15} integrator} \change{(see \citet{Yeager_2022_lifetimes} for additional detail on these data sets)}. This work focuses on the WHFast data set\change{, as} the WHFast integrator is significantly faster \change{than other symplectic integrators, enabling us} to integrate to the full 4.5 billion year age of the Solar System. To achieve parallelization, the 11.2 million ETA trajectories were placed into batches of 500, resulting in 22,400 parallel \change{simulated S}olar \change{S}ystems that were integrated forward \change{in time}. \change{Each} \change{simulated S}olar System \change{is} initialized with the Sun and eight planets \change{(i.e.,} Mercury, Venus, Earth-Moon, Mars, Jupiter, Saturn, Uranus, and Neptune\change{), with} initial planetary positions gathered from the JPL HORIZONS system\footnote{\change{\url{https://ssd.jpl.nasa.gov/horizons/}}} at an epoch of JD2459145.61625984 \citep{1996DPS....28.2504G}. We utilized the Catalyst, Quartz, and Ruby clusters managed by Livermore Computing at Lawrence Livermore National Laboratory \footnote{\url{https://hpc.llnl.gov/hardware/platforms/ruby}}.

\indent Each batch of asteroids were initialized with six orbital parameters selected from a random distribution, as described below. Each model was then integrated forward with \texttt{REBOUND}\footnote{\url{https://github.com/hannorein/rebound}}\citep{rebound}. The integration time step was set to one-day, and position and velocity information for each simulation was output every 1000 years. \change{Each} ETA orbit \change{was} initialized with a true longitude ($\theta$) randomly and uniformly distributed between Earth and L3 (on the side of Earth's orbit containing L4). The argument of pericenter and longitude of ascending node w\change{ere} randomly assigned values between 0 and $2\pi$, assuming a uniform distribution. The semi-major axis ($a$), eccentricity ($e>0$), and inclination ($i$) were sampled from a three-dimensional Gaussian with mean \change{($\mu$)} and covariance \change{($\Sigma$)} given in \Cref{eq:cov}.

\begin{equation}
    \label{eq:cov}
    \boldsymbol{\mu} = \begin{pmatrix} 
    \bar{a} \\
    \bar{e} \\
    \bar{i}
    \end{pmatrix} = \begin{pmatrix} 
    1\,\text{\change{AU}} \\
    0 \\
    0\degree{}
    \end{pmatrix} ~
    \boldsymbol{\Sigma} = \begin{pmatrix} 
    0.025\,\text{\change{AU}} & 0 & 0 \\
    0 & 0.075 & 0 \\
    0 & 0 & 15\degree{}   
    \end{pmatrix}.
\end{equation} 

\subsection{Modeling ETA intrinsic properties}
\label{sec:etaproperties}
\indent Every ETA in the MEGASIM is assigned an albedo, asteroid type, diameter, and absolute magnitude ($H$). The albedo of the ETAs are randomly sampled from the distribution curve of \change{near-Earth asteroids (}NEAs\change{)} determined by \citet{Wright_2016}\change{,} with a functional form given by \Cref{eq:albedo}\change{.} \move{\change{\citet{Wright_2016} fit two} Rayleigh distributions \change{to the NEA data to get the} albedo distribution\change{: one Rayleigh distribution} with \change{a} dark peak at \change{0.030 (}accounting for 25.3\% of the NEAs observed by WISE\change{),} and \change{one with a bright peak at 0.168 (}for the remaining 74.7\% of the NEAs\change{)}.}
\begin{equation}
    p(p_{v}) = {\exp}\left(\dfrac{-p^{2}_{v}}{2d^2}\right) \frac{f_{D} p_{v}}{d^2} + {\exp}\left(\dfrac{-p^{2}_{v}}{2b^2}\right)\, \frac{p_{v}}{b^2}\,(1 - f_{D})
    \label{eq:albedo}    
\end{equation} where $p$ is the probability density function, $p_{v}$ is the asteroid albedo, $f_{D} = 0.253$ \change{is} \move{the dark fraction}, $b = 0.168$ \change{is the} \move{bright peak,} and $d = 0.030$ \change{is the} \move{dark peak}.

\indent The `type' of the asteroid is either an `S' \change{(siliceous)} or `C' \change{(carbonaceous)} type and is determined via a second level of random sampling after the albedo is determined. The two Rayleigh functions that produce \Cref{eq:albedo} correspond to an `S' or `C' type asteroid. The height of each function is used to weight the random sample that determines whether an ETA is of `S' or `C' type\change{, and} the ETA type determines the correction magnitude for the final apparent brightness calculation.

\indent The asteroid diameter \change{($D_{\text{ast}}$)} is obtained from \change{magnitude (}$H$\change{)} and albedo ($p_v$) using \Cref{eq:HtoDconvert}. 
\begin{equation}
    D\textsubscript{ast} = \frac{1329000}{\sqrt{p_{v}}} \times 10^{-0.2 H}
    \label{eq:HtoDconvert}    
\end{equation}
$H$ was randomly sampled from an extended version of the distribution presented in \citet{GRANVIK2018181}, which covered $17<H<25$ with a break in the power law slope at $H=23$. To sample from the distribution, the number of asteroids at a given absolute magnitude was approximated using two linear functions, but we extrapolated the distribution to $H=28$.



\indent The apparent magnitude \change{($m_{V}$)} of \change{each} ETA is obtained via \Cref{eq:weighting}\change{.}
\begin{equation}
    \begin{aligned}
    m_{V} = H + 5 \log{}_{10} \left( \frac{R_{AS} R_{AE}}{R_{0}^2}\right) - 2.5 \log{}_{10} \left( q(\alpha)\right)
    \label{eq:weighting}
    \end{aligned}
\end{equation}where $R_{AS}$ is the distance between the Sun and the asteroid, $R_{AE}$ is the distance between the asteroid and the center of Earth, $R_{0}$ is a conversion factor of 1 AU to meters (the unit used for all distances), and $q(\alpha)$ is the phase integral of a diffuse reflecting sphere \change{given in} \Cref{eq:phaseintegral}\change{.}
\begin{equation}
    \begin{aligned}
    q(\alpha) = \frac{2}{3}\left( \left( 1-\frac{\alpha}{180\degree}\right) \cos{\alpha} + \frac{1}{\pi}\sin{\alpha} \right)
    \label{eq:phaseintegral}
    \end{aligned}
\end{equation} where $\alpha$ is the phase angle between the Sun\change{,} Asteroid\change{, and} Earth.

The magnitude of the Sun \change{(}$M_{Sun}$\change{)} is chosen to be 4.80 \move{for calculations in this paper}\change{, which corresponds to} the magnitude of the Sun in Johnson V-band \citep{Willmer_2018}. Corrections from the Johnson V\change{-}band to both the LSST and ZTF filters are calculated by \citet{DEMEO2009160}\change{,} \citet{chance2010improved}\change{, and} \citet{chesley2017projected}, the values of which are provided in \Cref{table:colorcorrection}.

\begin{deluxetable}{|c|cccccc|}[htb]
    \tabletypesize{\small}
    \tablecaption{{\change{LSST Color Corrections}}
    \label{table:colorcorrection}}
    \tablehead{
        \multicolumn{1}{|l|}{\textbf{Class}} & 
        \multicolumn{1}{c}{\textbf{V-u}} &
        \multicolumn{1}{c}{\textbf{V-g}} &
        \multicolumn{1}{c}{\textbf{V-r}} &
        \multicolumn{1}{c}{\textbf{V-i}} &
        \multicolumn{1}{c}{\textbf{V-z}} &
        \multicolumn{1}{c|}{\textbf{V-y}}
    }
    \startdata
        C & -1.614 & -0.302 & 0.172 & 0.291 & 0.298 & 0.303 \\
        S & -1.927 & -0.395 & 0.255 & 0.455 & 0.401 & 0.406
    \enddata
    \tablecomments{\change{Johnson V-band to LSST filter system color corrections.}}
\end{deluxetable}

\subsection{\move{The LSST and ZTF \change{S}urveys}}
\indent \move{\change{We use the} LSST v2.2 baseline and twilight \texttt{OpSim} runs\footnote{\url{https://www.lsst.org/scientists/simulations/opsim}} \change{(`lsst\_no\_twilight\_neo\_v2.2' and `twi\_neo\_repeat4\_riz\_v2.2\_10yrs', respectively),} as well as all ZTF observations spanning dates 2018-03-17 to 2022-06-01 \citep{lsstopsim,bellm2018zwicky}\change{, as the survey data for our analysis (see} section \ref{sec:results}\change{)}. \move{The number of observations in the LSST baseline \texttt{OpSim} was 2,259,570 and twilight 2,078,065.} \change{We filter down} the ZTF data set to only include 30-second exposures taken by the Palomar P48 observing system, \change{and we use} both public \change{survey data} and private \change{institutional} partnership observation\change{al data}. The \change{resulting} dataset consists of 706,459 30-second observations.}

\subsection{\new{Estimating \change{S}urvey \change{L}osses}}
\indent \new{To be considered a \move{detection, one of two conditions must be satisfied: 1)} three pairs of observations \change{(}\move{at minimum}\change{)} are required within a month-long rolling window\change{, or 2)} two observations \change{are required} within one hour of one another. These conditions are reasonable for orbit fitting between observations \citep[\change{e.g.,}][]{jedicke2006pan,denneau2009asteroid} and are to be used \change{by} LSST \citep{2018Icar..303..181J}.} \new{Survey detections \change{can be} lost from non-ideal conditions such as observation linking, trailing losses from streaking across \change{a detector (i.e.,} the CCD\change{)}, and chip gaps between pixels \change{on the detector}. Loss of detection due to the gaps between CCD pixels is accounted for via random sampling. A `fill factor'\change{, i.e.,} the ratio of total pixel area to total CCD area, is used to approximate the likelihood that an asteroid would fall between pixels. The fill factor for LSST is 0.877\change{,} and for ZTF \change{is} 0.875 \citep{2018Icar..303..181J,bellm2018zwicky}}

\begin{deluxetable}{|l|c|c|c|}[htb]
    \tabletypesize{\small}
    \tablecaption{{\change{Simulation Parameters}}
    \label{table:etc_values}}
    \tablehead{
        \multicolumn{1}{|l|}{\textbf{Parameter}} & 
        \multicolumn{1}{c|}{\textbf{Units}} &
        \multicolumn{1}{c|}{\textbf{LSST}} &
        \multicolumn{1}{c|}{\textbf{ZTF}}
    }
    \startdata
         Exposure Time & seconds & 30 & 30 \\ \hline
         Pixel Scale & $\text{arcsec}$ $\text{pixel}^{-1}$ & 0.2 & 1.0 \\ \hline
         Read Noise & $\text{e}^{-}$ $\text{pixel}^{-1}$ & 8.8 & 8.0 \\ \hline
         Dark Current & $\text{e}^{-}$ $\text{pixel}^{-1}$ $\text{s}^{-1}$ & 0.2 & 1.0 \\ \hline
         Gain & $\text{e}^{-}$ $\text{ADU}^{-1}$ & 2.3 & 6.2 \\ \hline
         Aperture$^{*}$ & meters & 6.423 & 0.4165 \\ \hline
         PSF FWHM & arcsec & 0.87 & 2.0 \\ \hline
         Sky Brightness & AB mag $\text{arcsec}^{-2}$ & u=22.8 & g=20.75 \\
          & & g=22.3 & r=20.23 \\
          & & r=21.2 & i=19.57 \\
          & & i=20.5 & \\
          & & z=19.6 & \\
          & & Y=18.6 & \\
    \enddata
    \tablecomments{\new{$^{*}$Mean effective diameter of light collecting region. LSST is assumed to have a perfectly circular field-of-view, while ZTF is assumed to be a perfect square.}}
\end{deluxetable} 

\indent \change{To estimate trailing losses \change{(i.e., the loss in flux from streaking over the detector)} due to asteroid streaking over an exposure, we run an exposure time calculator using simulation parameter values from Table \ref{table:etc_values}, to determine the limiting magnitude for each survey filter (i.e., u-, g-, r-, i-, z-, and Y-band for LSST, and g-, r-, and i-band for ZTF), over the set of possible angular rates (75 angular rates equally spaced between $0.1$ and $1.0$ arcseconds per second). We assume that the LSST and ZTF surveys are operating in their traditional pipeline settings, in which observations are fit using a PSF (even though the ETA is streaking across the detector)\footnote{\change{Worth noting is that you can get a significant increase in asteroid detectability by fitting images with a streak profile.}}. For both surveys, the detection limit is set to $5\sigma$, and the exposure time is 30 seconds. LSST baseline atmospheric throughput (with aerosols)\footnote{\change{atmos\_std.dat from {\url{https://github.com/lsst/throughputs}}}} is used for both the LSST and ZTF simulations. ZTF simulations use the Palomar ZTF total throughput data (i.e., filter + sensor) for the g-, r-, and i- bands\footnote{\change{From \url{http://svo2.cab.inta-csic.es/svo/theory/fps3}}}, and LSST simulations use total throughput for the u-, g-, r-, i-, z-, and Y-bands\footnote{\change{`filter\_$\ast$.dat' from \url{https://github.com/lsst/throughputs}}}. For LSST we assume a pixel scale of 0.2 arcseconds per second, read noise of 8.8 electrons per pixel, dark current of 0.2 electrons per pixel per second, gain of 2.3 photoelectrons per count, effective telescope diameter of 6.423 meters, and a point-spread function (PSF) full-width half-max (FWHM) of 0.87 arcseconds (all of which are publicly reported\footnote{\change{\url{https://www.lsst.org/scientists/keynumbers}}}). The corresponding values for ZTF are 1.0, 8.0, 1.0, 6.2, 0.4165, and 2.0, respectively \citep{Dekany:2020}. LSST sky brightness values are taken from the reported sky count numbers at the time of the simulations\footnote{\change{\url{https://smtn-002.lsst.io} (values are occasionally updated).}}. ZTF sky brightness values are calculated by convolving the 50th percentile dark sky spectral energy distribution as a function of wavelength (from the Gemini Near-Infrared Instrument) with each ZTF passband (recalibrated using an AB magnitude zero-point), to determine the median magnitude of the sky brightness per filter \footnote{\change{Note that we use the same dark sky values for the LSST twilight survey and baseline survey}}. We then determine whether our simulated asteroids would be detected by the given survey or not by cutting out any ETA whose observation falls above the limiting magnitude for that angular rate and filter combination.}

\subsection{\move{Translating MEGASIM \change{O}rbits to \change{S}urvey \change{P}ointings}}
\label{sec:surveypointings}
\indent \move{Of the 11.2 million MEGASIM initialized orbits, only those stable for \change{a minimum of} 15\change{,000} years (\move{$\sim$368,287 ETA orbits}) were propagated through the survey simulations. This propagation \change{uses} a maximum time step of 30 seconds\change{, and is} integrated to the exact time of each \change{simulated} LSST and ZTF observation \change{time}\footnote{\change{The time of each simulated observation comes from \texttt{OpSim} for our simulated LSST surveys, and real observation times are taken for our simulated ZTF survey}}. \change{If \move{observation times are shorter than the 30 second timestep, then we integrate to the exact timestep of the exposure.}} To simulate how many MEGASIM ETAs could be detected in the LSST and ZTF surveys, we propagate the simulation \change{forwards or backwards} from  \change{$t_0$=JD2459145.61625984 (the time that our surveys are initialized at)} to the start date of the respective surveys; for LSST, forward propagation was done to sync with the start date of the LSST \texttt{OpSim}, and backward propagation was done to sync the \change{S}olar \change{S}ystem with the beginning of the ZTF survey. The shifting of initialization resulted in on-sky errors of the order $10^{-6}$\degree{} in right ascension (R\change{.}A\change{.}) and $6\times10^{-7}$\degree{} in declination  (Dec.). A maximum error \change{in pointing} of \change{2.25} arcseconds in R.A. and Dec. occurred over the course of the entire simulation. Considering the effective field\change{-}of\change{-}view diameters of 1.75\degree{} for LSST and 3.4\degree{} for ZTF, the deviations between the simulated and true positions do not significantly affect results.}

\indent \new{Once the MEGASIM \change{simulated S}olar \change{S}ystem reaches the timestamp of a given survey \change{pointing}, a check is done to determine which ETAs are at an R.A. and Dec. within the observations field-of-view. To simplify this check, a circle is used to approximate the field-of-view of LSST and a square for ZTF. The effective circle for LSST is taken to be of radius 1.75\degree{}, and for ZTF, the square has side lengths of 3.428\degree{} and an area of 47 sq degrees. ETAs found within a survey's \change{field-of-view} are then cataloged. A further check is then done to determine if the ETA in the field\change{-}of\change{-}view is bright enough to be detected by the telescope. If the ETA apparent magnitude is brighter than the 5$\sigma$ sensitivity for that exposure, it is cataloged as a detection.}

\subsection{\move{Generating \change{D}etection \change{S}tatistics}}
\indent \move{To place statistical weights on ETA detections in the LSST and ZTF surveys\change{, d}etections in each survey simulation were recalculated by assigning all ETAs a new \change{magnitude ($H$)}, diameter \change{($D_{\text{ast}}$)}, and albedo \change{($p_v$)}. \change{N}ew ETA properties are obtained via resampling the $H$ and $p_v$ distributions laid out in \S \ref{sec:etaproperties} \change{, enabling} ETA resampling and detection \change{determination} in post without needing to re-simulate all of the ETA trajectories. \change{This} resampling was completed 1000 times for each LSST \texttt{Opsim} and ZTF pointing. In each step \change{of the resampling}, a unique detection list was created following the procedure described above.}

\section{Distributions of ETAs}
\label{sec:maps}

\subsection{MEGASIM coordinates}
\label{sec:coordinates}
\indent The Cartesian coordinates of the MEGASIM are aligned with aspects of the \change{S}olar \change{S}ystem as it was on $t_0$=JD2459145.61625984. The X-axis is aligned with the vernal equinox at simulation time $t$ = $t_0$, with both the X\change{-} and Y-axis in the plane of the ecliptic. The Z-axis is aligned with the ecliptic north pole. We define a rotating ecliptic spherical coordinate system centered on Earth. The Sun's latitude and longitude are always at the origin in this rotating frame, and the idealized L4 point is at (0\degree, 60\degree). This simplifies the analysis by allowing us to simply stack parallel simulations with one another\change{, while} also satisf\change{ying} the goal of forecasting ETA \move{observability}.

\begin{figure}
    \centering
    \includegraphics[width=0.475\textwidth]{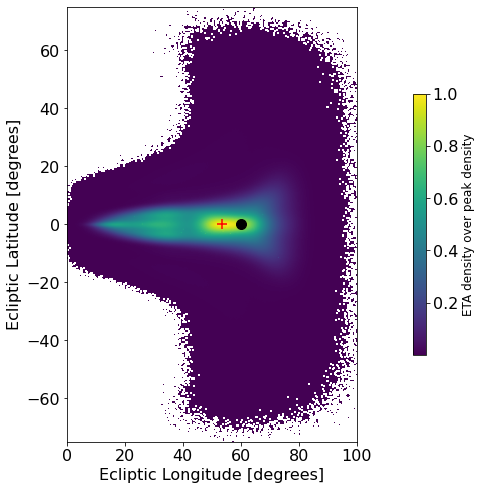}
    \caption{Stacked map of ETA \change{positions} from the WHFast integrated simulation \change{(including} all ETAs\change{, with a point at} every 100 kyr\change{,} up to 1 Gyr. The ecliptic coordinates are explained in Section \ref{sec:coordinates}. The Sun is at the origin, the black dot is L4 at (0\degree{},60\degree{}), and the red cross indicates the location of highest density, which is 7\degree{} offset from L4 \change{Sun-ward}. The bin \change{counts} are scaled to the bin with the highest density.}
    \label{fig:fullstacklatlon}
\end{figure}

\subsection{Stacked \change{D}istribution \change{M}aps}
\label{sec:stackeddist}

\indent \Cref{fig:fullstacklatlon} shows the number density of all ETA positions stacked in 100 kyr intervals\change{,} up to 1 Gyr. As ETAs are removed from the MEGASIM due to instability, they no longer contribute to these maps. Unstable orbits are more numerous\change{,} but offer significantly shorter trajectories than the few orbits that are stable on long timescales. Th\change{is} figure is presented in the ecliptic rotating coordinate system discussed above. The black circle is the L4 point (0\degree,60\degree), and the red cross is the bin with the highest number of ETAs. The peak density is 7\degree{} Sun-ward\change{,} offset from L4. The\change{se} results are largely consistent with a similar figure from \citet{2000Icar..145...33W}\change{,} but with a much higher resolution. The stacked 2\change{D} histograms reveal the large on-sky area over which ETAs can travel, but for the most part, ETAs are found within 10\degree{} of the ecliptic. There are several bands of higher and lower stability\change{, which} are \change{more} evident in later figures with different coordinates presented. \change{Worth noting} are \change{the} ridges and gaps above and below the ecliptic plane \change{at} longitudes between $\sim$20\degree{} and $\sim$50\degree{}, as well as wings that flare away from the ecliptic at longitudes beyond L4\change{, as these features have not been resolved in previous simulations}.

\begin{figure}
    \centering
    \includegraphics[width=0.375\textwidth]{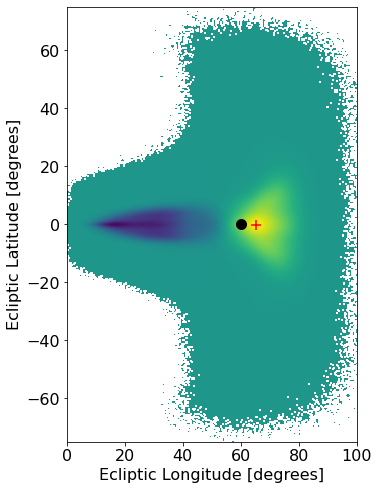}
    \caption{A difference map produced by weighting ETA trajectories by apparent magnitude, binning into a 2\change{D} histogram, normalizing all bins such that the heights exist in the range of 0 to 1\change{,} then subtracting off the bin heights \change{from} \Cref{fig:fullstacklatlon}. The yellow \change{shades} indicate where ETA apparent magnitude weighting is positive, green is near zero, and purple is negative. The black circle is the L4 point, and the red cross shows the location of the highest deviation from the density plot in \Cref{fig:fullstacklatlon}. The peak of the flux-weighted density map is located 6.5\degree{} Sun-ward from L4.}
    \label{fig:differencelatlon}
\end{figure}

\begin{figure}
    \centering
    \includegraphics[width=0.475\textwidth]{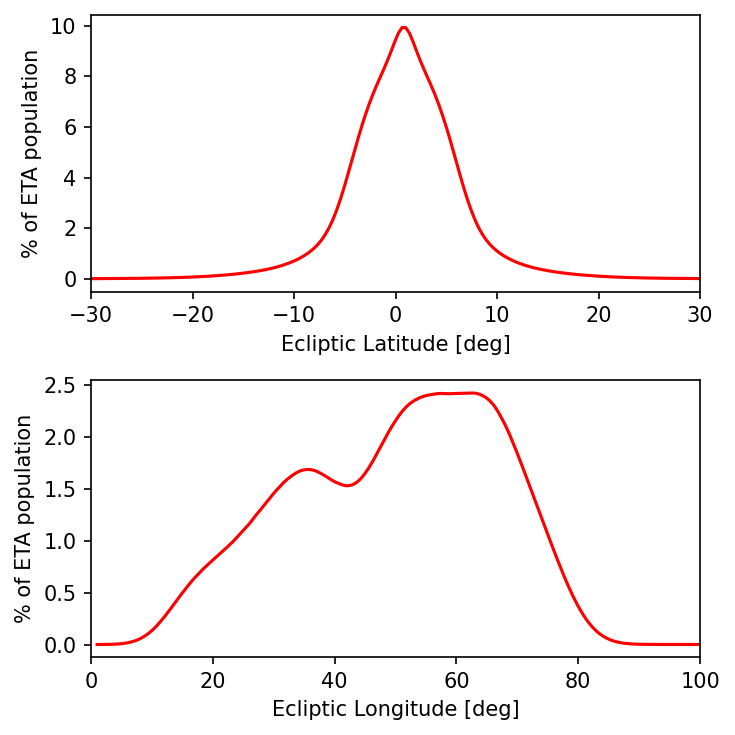}
    \caption{One degree slices of \cref{fig:fullstacklatlon} in both latitude and longitude. The y-ax\change{e}s \change{are} the percent of ETA trajectories found for a given one degree thick cross section in \cref{fig:fullstacklatlon}.}
    \label{fig:1dcuts}
\end{figure}

\indent Weighting the stacked ETA trajectories \change{from} \Cref{fig:fullstacklatlon} by their apparent magnitude has only a minor effect in determining where they may be observed. If the ETA on-sky positions are weighted by their flux, the location of the peak density is only shifted by about 0.5\degree{}. One\change{-}dimensional cuts \change{on} ecliptic longitude and latitude of \Cref{fig:fullstacklatlon} are shown in \Cref{fig:1dcuts}. The notable features are a steep fall off as the ecliptic latitude increases\change{,} a local minimum at $\sim$40\degree{}\change{,} and local maximum $\sim$35\degree{} ecliptic longitude. In \Cref{fig:differencelatlon} we show the difference obtained from the ETA trajectories by \change{weighting observations by} brightness \change{(i.e., apparent magnitude)}, binning \change{data into a 2D histogram}, and \change{renormalizing bin values into} the range of 0 to 1 \change{and taking the difference (see \Cref{fig:fullstacklatlon})}. ETAs interior to L4 are generally fainter than the density map would suggest (hence the blue \change{patch} interior to the L4 point). This is due to the majority of ETAs in this portion of the sky existing at fairly far distances from Earth, resulting in lower apparent magnitude. Those found in the region exterior to L4 from the Sun tend to be close to their nearest approach to Earth on their trajectory. The phase angle in the brightest region does result in a reduction of apparent magnitude, but the effect is sub-dominate to the proximity effect with Earth. For observing considerations, the region behind L4 is much easier to observe due to the greater solar elongation \change{(}compared \change{to} trying to observe the generally fainter ETAs interior on the sky to L4\change{)}.

\begin{figure}
    \centering
    \includegraphics[width=0.475\textwidth]{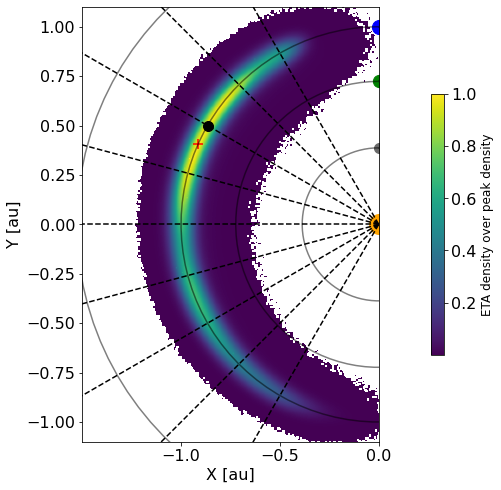}
    \caption{A top-down view of the \change{S}olar \change{S}ystem. The Sun is at (0,0), and the concentric black circles starting nearest the Sun indicate the approximate orbits of Mercury, Venus, Earth, and Mars. Radial dashed lines are provided for angular reference at 30\degree{}, 45\degree{}, 60\degree{}, 75\degree{}, 90\degree{}, 105\degree{}, 120\degree{}, 135\degree{}, \change{and} 150\degree{} measured counter-clockwise from the location of Earth in the figure. Binning is \change{done following} the same process as \Cref{fig:fullstacklatlon}}
    \label{fig:fullstackxy}
\end{figure}

\indent \Cref{fig:fullstackxy} shows the projected density of stacked ETA positions in the ecliptic plane\change{, and} reveals that ETA trajectories extend into the \change{S}olar \change{S}ystem as far as the orbit of Venus\change{,} span\change{ning} the full range of longitudes from Earth to L3. As one would expect, the highest ETA densities are found around L4, but slightly counterclockwise from L4, which agrees with the location indicated in \Cref{fig:fullstacklatlon}. Measuring an angle counter-clockwise from Earth \change{(the blue dot} at the top of \Cref{fig:fullstackxy}\change{)}, the highest density \change{of ETAs} is located about 5\degree{} \change{from L4, towards the} Earth. ETA density around the peak remains within 10\% of the peak as close as 45\degree{} out to about 80\degree{} from Earth. From 90\degree{} to 105\degree{}, there is a region of reduced ETA density before another local maximum in ETA density found near 115-120\degree{} from Earth.

\subsection{Distribution of \change{B}illion \change{Y}ear \change{S}table ETAs}
\label{sec:gyretas}

\begin{figure}
    \centering
    \includegraphics[width=0.475\textwidth]{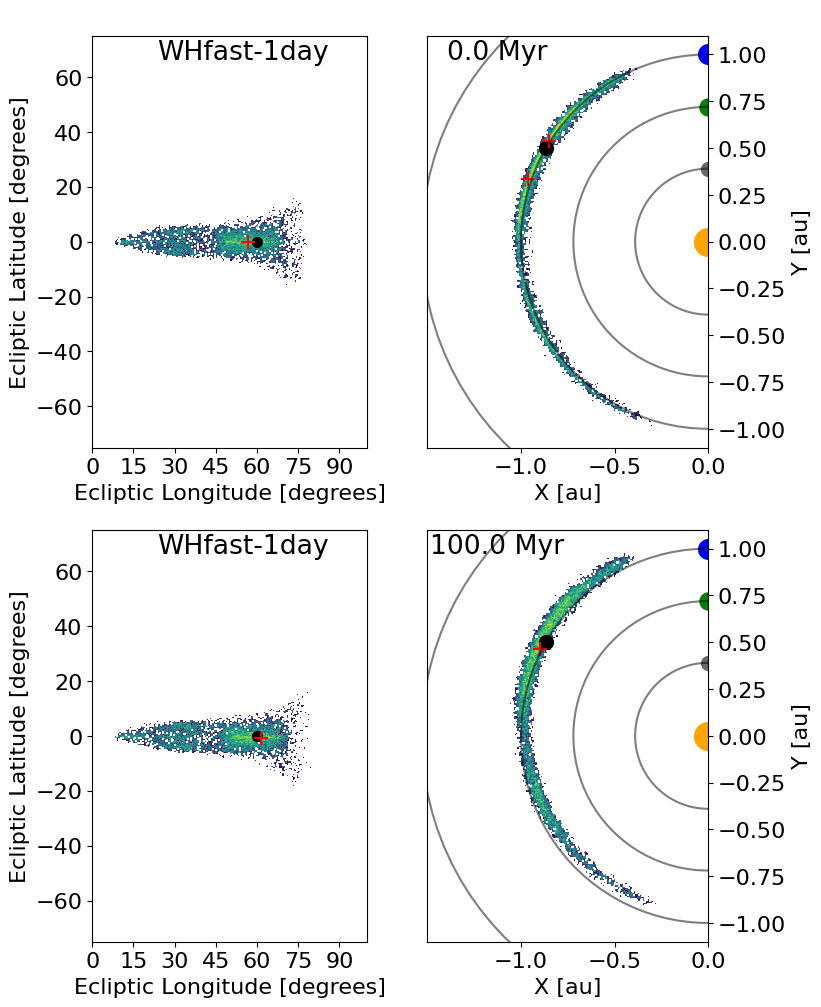}
    \caption{The distributions of only those ETAs that survive at least 1 Gyr of \change{the} simulation. \change{The top panels represent the initial positions of all ETAs that are stable for at least 1 Gyr, while the bottom panels show 100 Myr after those initial positions}. In each panel, the black dot is the approximate location of L4, the red cross is the bin with the highest Gyr ETA density, and the circles are at the semi-major axis distances of Mercury, Venus, Earth, and Mars. The location of the highest Gyr ETA density oscillates between 45\degree{}-70\degree{} latitude on-sky and in the X-Y plane, travels as close as 45\degree{} leading Earth to as far as 90\degree.}
    \label{fig:Gyrstabledist}
\end{figure}

\indent \Cref{fig:Gyrstabledist} shows single snapshots in time\change{, for only those} ETAs that have survived at least 1 Gyr of the MEGASIM integration. The top panels are the initial positions of all billion year (Gyr) stable ETAs, and the bottom panels are 100 Myr \change{after the initial positions}. To state the most obvious first, Gyr-stable ETAs are of low inclination and do not extend beyond 10\degree ecliptic latitude. The spatial distribution, unsurprisingly, does not span far from that of Earth's orbit. The same density ridges and gaps seen in \Cref{fig:fullstackxy} persist in the Gyr stable orbits. The tail of the ETA distribution is within the drawn circular Earth orbit at 100 Myr, \change{shown} in the bottom right panel of \Cref{fig:Gyrstabledist}. Other snapshots in time reveal the distribution of ETAs oscillates in and outside of the drawn circular orbit of Earth\change{,} at a frequency higher than the simulation output of 1000 years. This indicates that there is a bulk forcing of the ETA population throughout their libration trajectories \change{i.e., all ETAs are moving together as a relatively cohesive unit, and that distribution wobbles around together with a wave like pattern, as they move with Earth in its orbit)}.

\section{Results: The \change{D}etection \change{Prediction} of ETAs in ZTF and LSST \change{S}urveys}
\label{sec:results}

\subsection{Detection \change{S}tatistics and \change{C}umulative \change{C}ounts}

\begin{figure}
    \centering
    \includegraphics[width=0.475\textwidth]{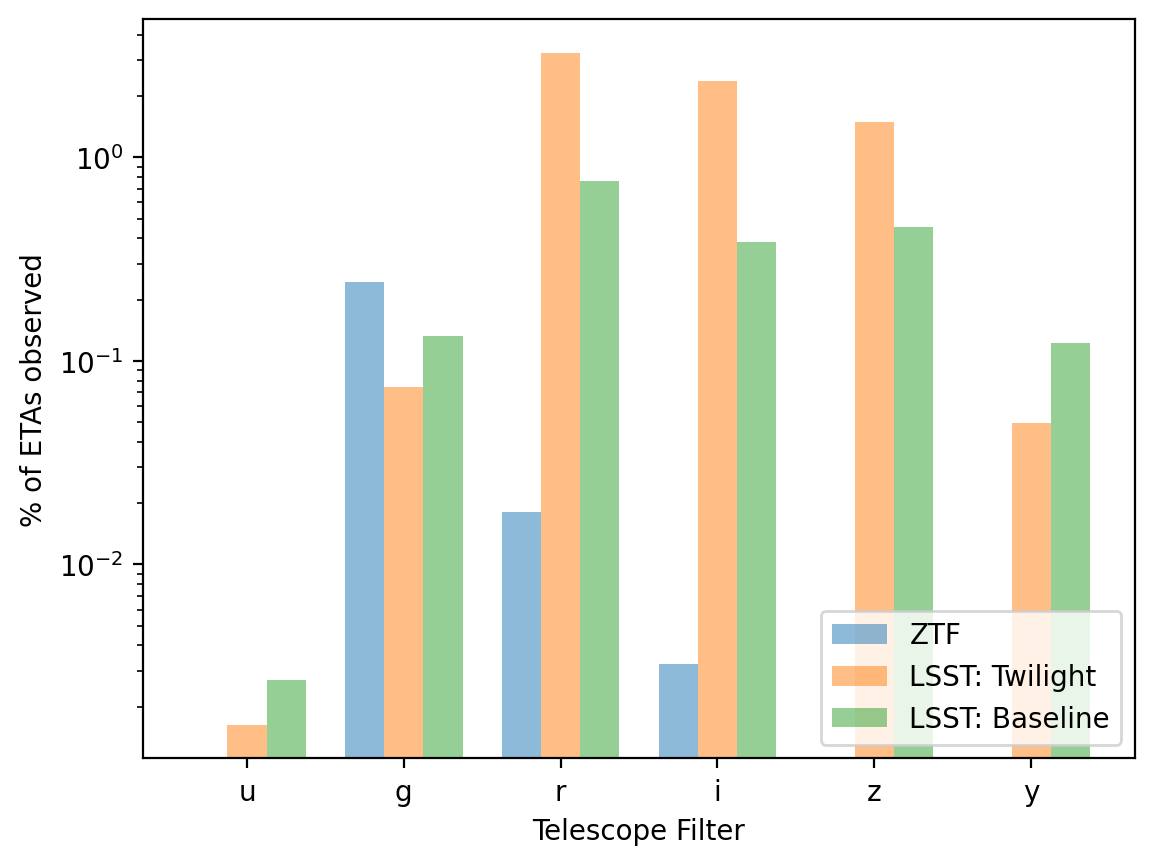}
    \caption{Detected ETAs by survey \change{and} filter type. Bins are normalized to the total propagated 1 Myr stable ETA trajectories (368,287 \change{ETAs}).}
    \label{fig:filterdetections}
\end{figure}

\indent \change{\Cref{fig:filterdetections} shows the percentage of ETAs by filter type, split between the ZTF survey (blue), LSST twilight survey (yellow), and LSST baseline survey (green).} The r-band filter for LSST and g-band filter for ZTF are the most successful at detecting ETAs\change{, and} ZTF's g-band detects more ETAs than LSST's g-band. Of the 706,459 ZTF pointings used\change{,} 38.57\% \change{(}272,541) were in g-band, 56.99\% \change{(}402,632) \change{in} r-band, and 4.43\% \change{(}31,286) in i-band. For the LSST baseline \texttt{OpSim} \change{results}, 18.67\% \change{(}387,928), 21.71\% \change{(}451,222), 22.66\% \change{(}470,990), 19.61\% \change{(}407,434), 10.96\% \change{(}227,705), and 6.39\% \change{(}132,786) \change{pointings were in the u-, g-, r-, i-, z-, and y-bands,} respectively \change{(and the twilight simulation follows roughly the same distribution)}. The ratio of the number of g-band and r-band observations by LSST \change{are} of unity but g-band provides an order of magnitude fewer detections than ZTF.

\begin{figure}
    \centering
    \includegraphics[width=0.475\textwidth]{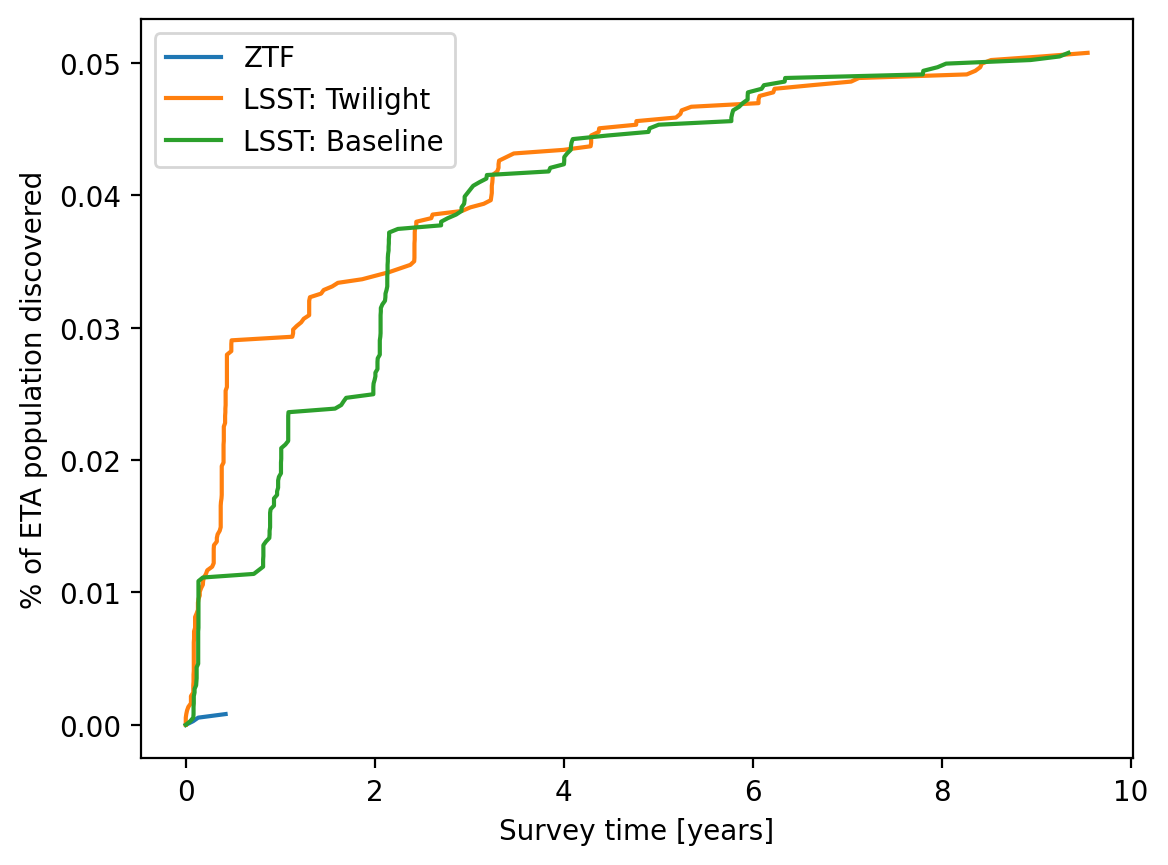}
    \caption{The cumulative \change{percentage} of \change{unique} ETA\change{s} detect\change{ed} \change{(from a population of 368,287 ETAs)} made by each survey for the fiducial simulation.}
    \label{fig:cumulativepercent}
\end{figure}

\begin{figure}
    \centering
    \includegraphics[width=0.475\textwidth]{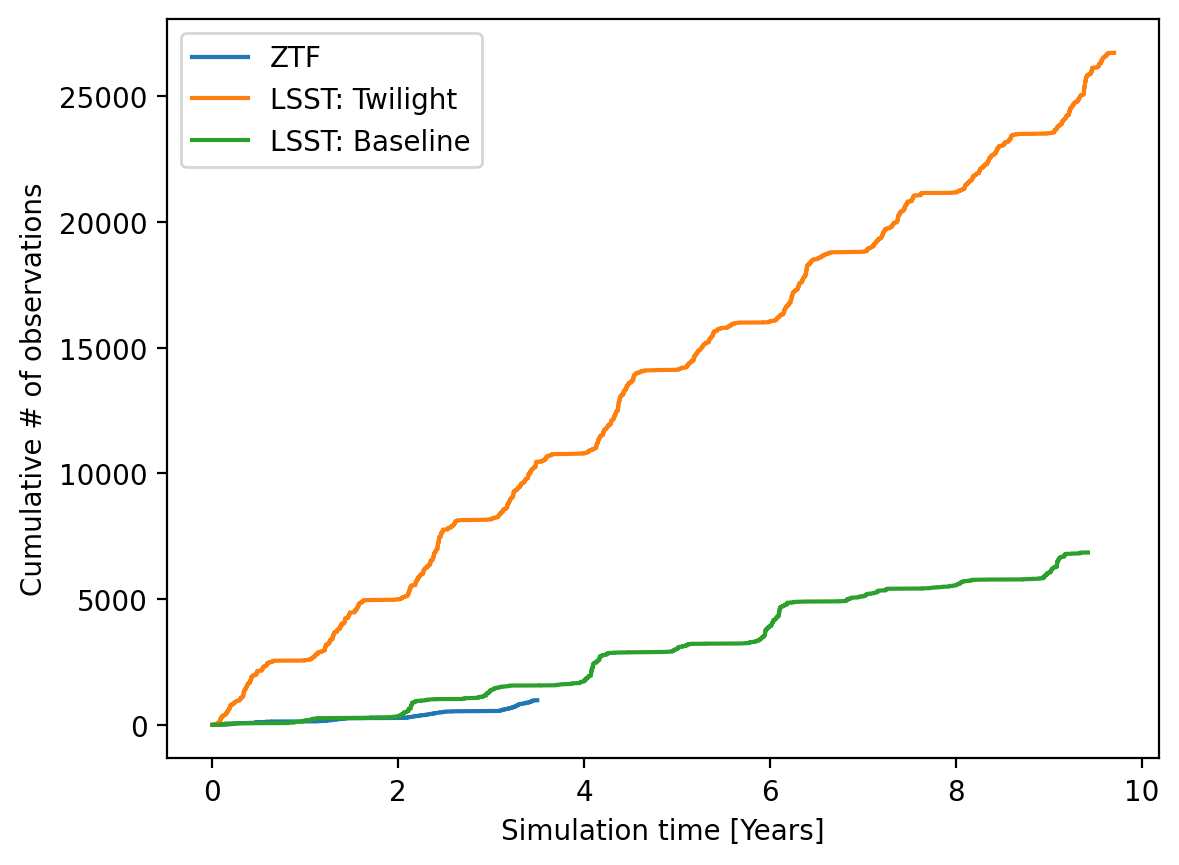}
    \caption{\change{The cumulative number of ETA detections made by each survey for the fiducial simulation.}}
    \label{fig:cumulativenumber}
\end{figure}

\indent The cumulative number of ETA detections for ZTF and the LSST \texttt{OpSim} \change{results are} provided in \Cref{fig:cumulativenumber}\change{, and the cumulative percent of the ETA population that is detected over the survey time are shown in \Cref{fig:cumulativepercent}.} The rate ZTF detects ETAs is roughly the same as the baseline LSST \texttt{OpSim} \change{result}. \change{The simulated t}wilight LSST \change{survey} captures approximately four times the number of ETA detections as \change{the} LSST baseline or ZTF.

\begin{figure}
    \centering
    \includegraphics[width=0.475\textwidth]{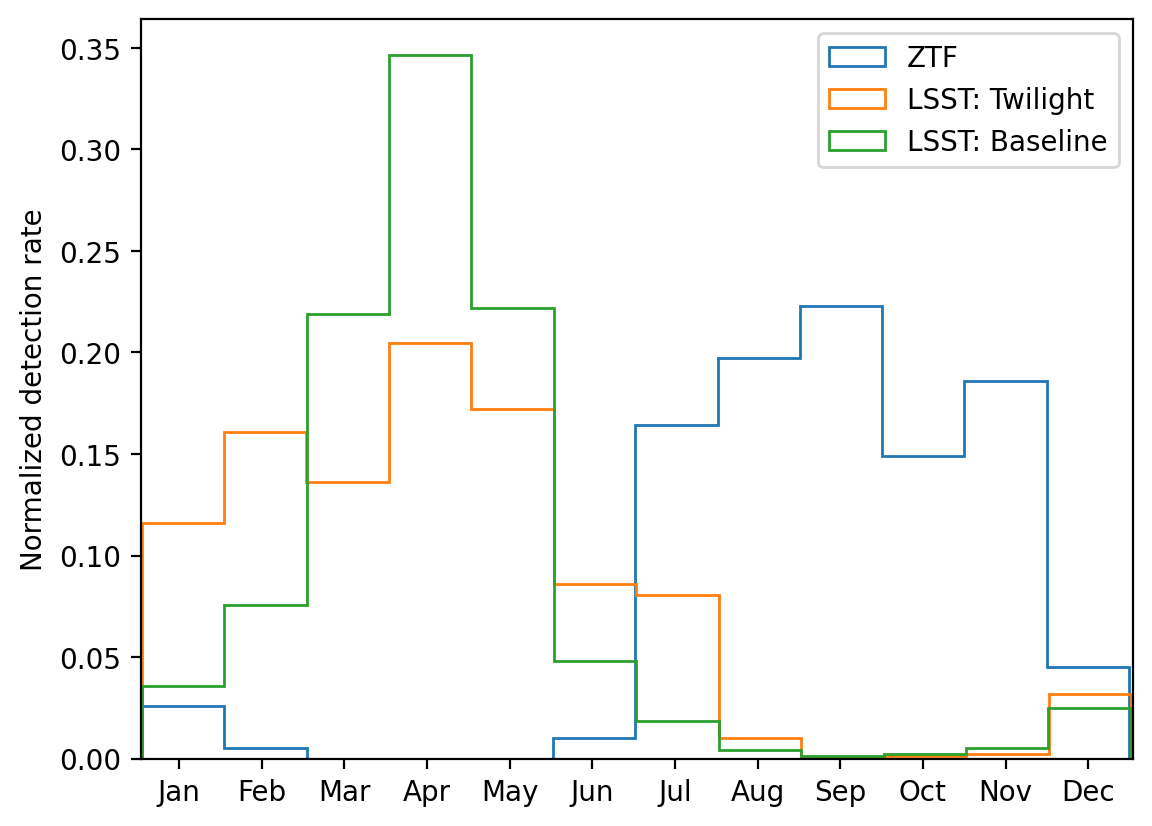}
    \caption{The normalized number of detections by month\change{,} made by each survey.}
    \label{fig:detectionspermonth}
    
\end{figure}

In \Cref{fig:detectionspermonth}, we show the relative detection rate for each survey by month. Variations occur in the detection rate of ETAs as the time a telescope can be on target toward L4 varies over the year \citet{WHITELEY1998154}. Beyond the variable detection rate, ZTF finds \change{the} most ETAs \change{between} approximately June \change{and} December \change{(with} \move{nearly the same number of ETAs} \change{detected between} \move{August} \change{and} \move{November} \change{)}, whereas LSST makes \change{the most} ETA detections from \change{approximately} January to June \change{(with} \move{the most ETAs} \change{detected} \move{in the month of April}\change{)}. These differences are due to the different hemispheres of each observatory, which changes the time of year L4 is visible.

\begin{figure}
    \centering
    \includegraphics[width=0.475\textwidth]{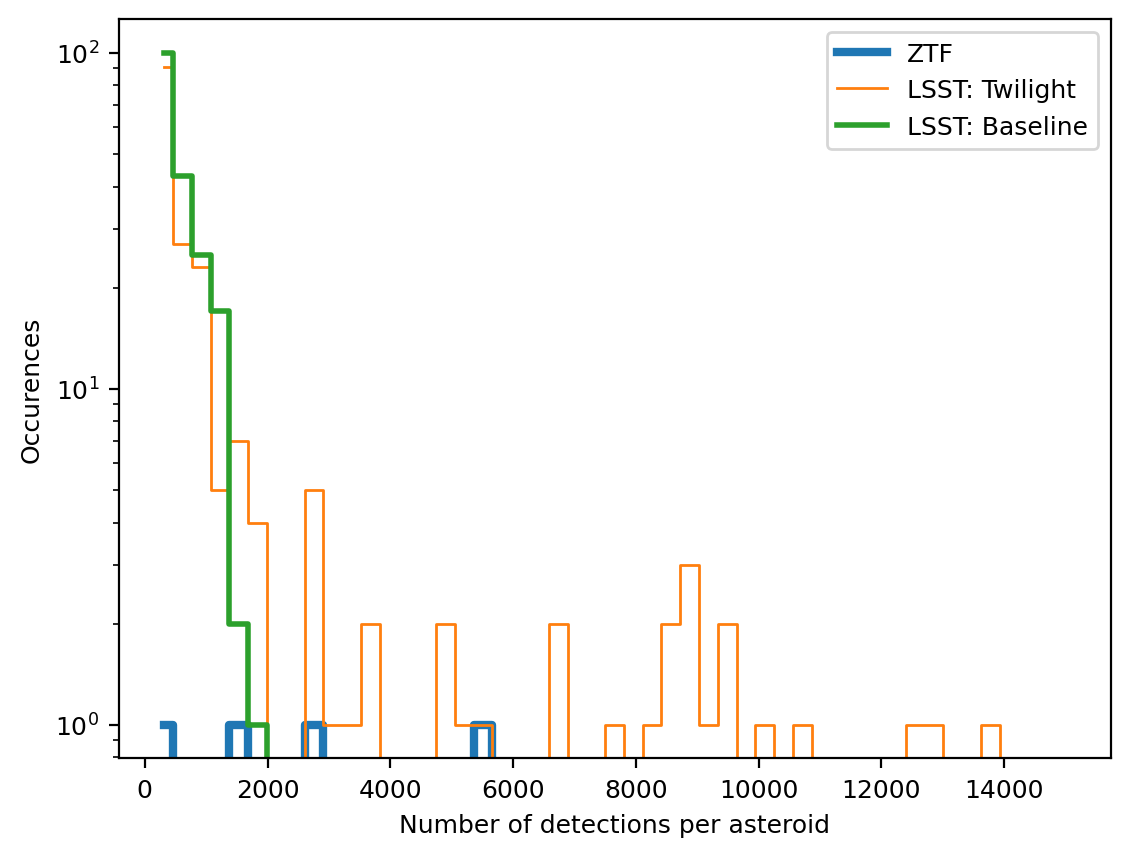}
    \caption{Asteroid recovery by each survey.}
    \label{fig:detectionsperasteroid}
\end{figure}

\indent The number of distinct detections per asteroid\change{, per} survey\change{,} is shown in \Cref{fig:detectionsperasteroid}. There is a stark difference between the two LSST surveys. This is because the twilight survey observes lower Solar elongations nightly, which is capable of recovering the same asteroid thousands of times as it slowly librates through the twilight survey footprint. The ZTF survey also has a twilight survey, which explains the few instances of an ETA being detected thousands of times; however, it is far less sensitive, so this only works for a few bright ETAs. The baseline LSST survey recovers many asteroids numerous times \change{(}\move{as many as 2000 times}\change{, which} \move{is far more than typical} \change{NEAs} \move{are recovered}\change{)}, but not nearly as many as the LSST \texttt{OpSim} twilight survey \change{results}. \citet{2018Icar..303..181J} estimate that typical NEOs are recovered $\sim$100 times.

\begin{figure}
    \centering
    \includegraphics[width=0.475\textwidth]{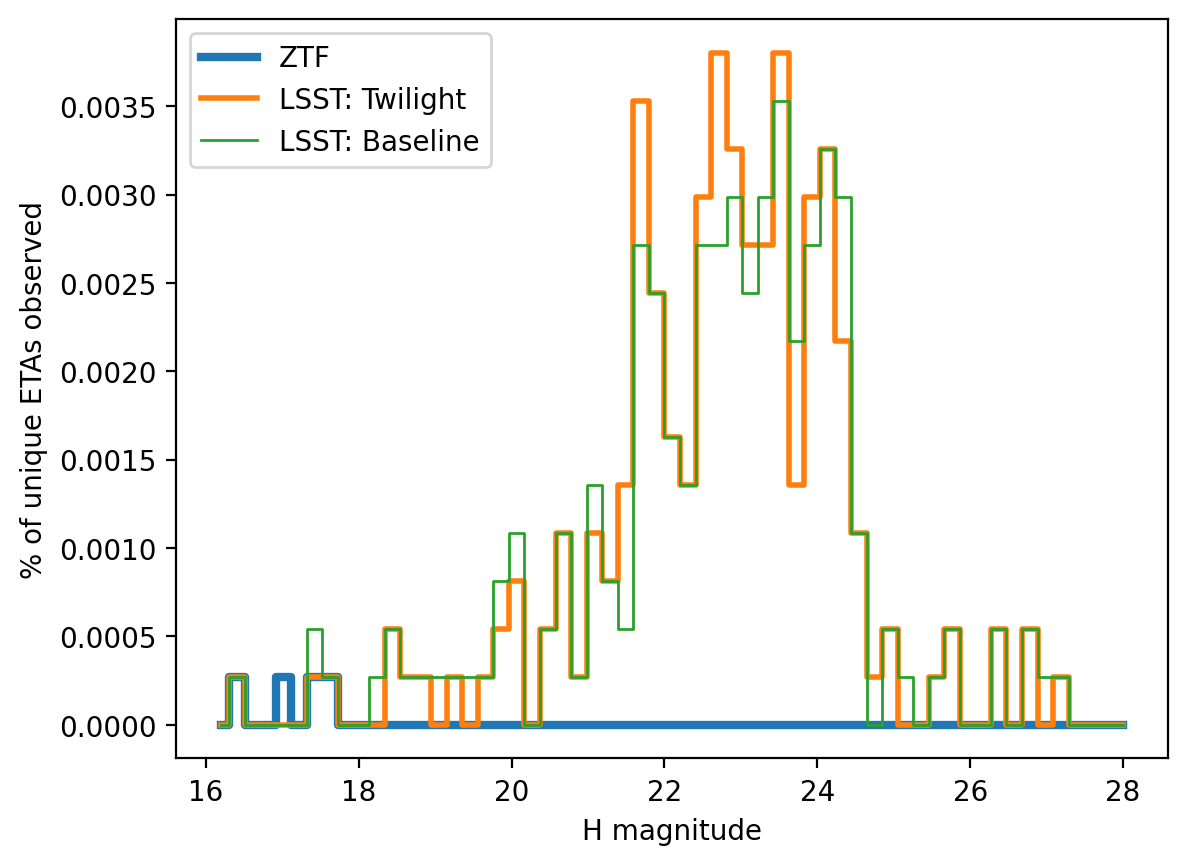}
    \caption{Detected ETA \change{percentage as a function of} $H$. Bins are normalized to the total propagated 1 Myr stable ETA trajectories (368,287 \change{ETAs}).}
    \label{fig:h_magnitude}
\end{figure}

\indent \Cref{fig:h_magnitude} shows the distribution of $H$ for \change{unique} detected ETAs. Unique detections by ZTF are few and do not occur for $H>20.5$. For LSST, there is very little difference in the distribution of $H$ when comparing the baseline and twilight \texttt{OpSim} surveys. LSST makes the bulk of detections in the range $21<H<25$, though some detections are made up to \change{$H<27.5$}. This suggests that ETAs come close enough to Earth to be detected by LSST \change{at} extremely faint magnitudes, which is a statement of the Rubin Observatory's sensitivity \change{and observing pattern}. 

\begin{figure}
    \centering
    \includegraphics[width=0.475\textwidth]{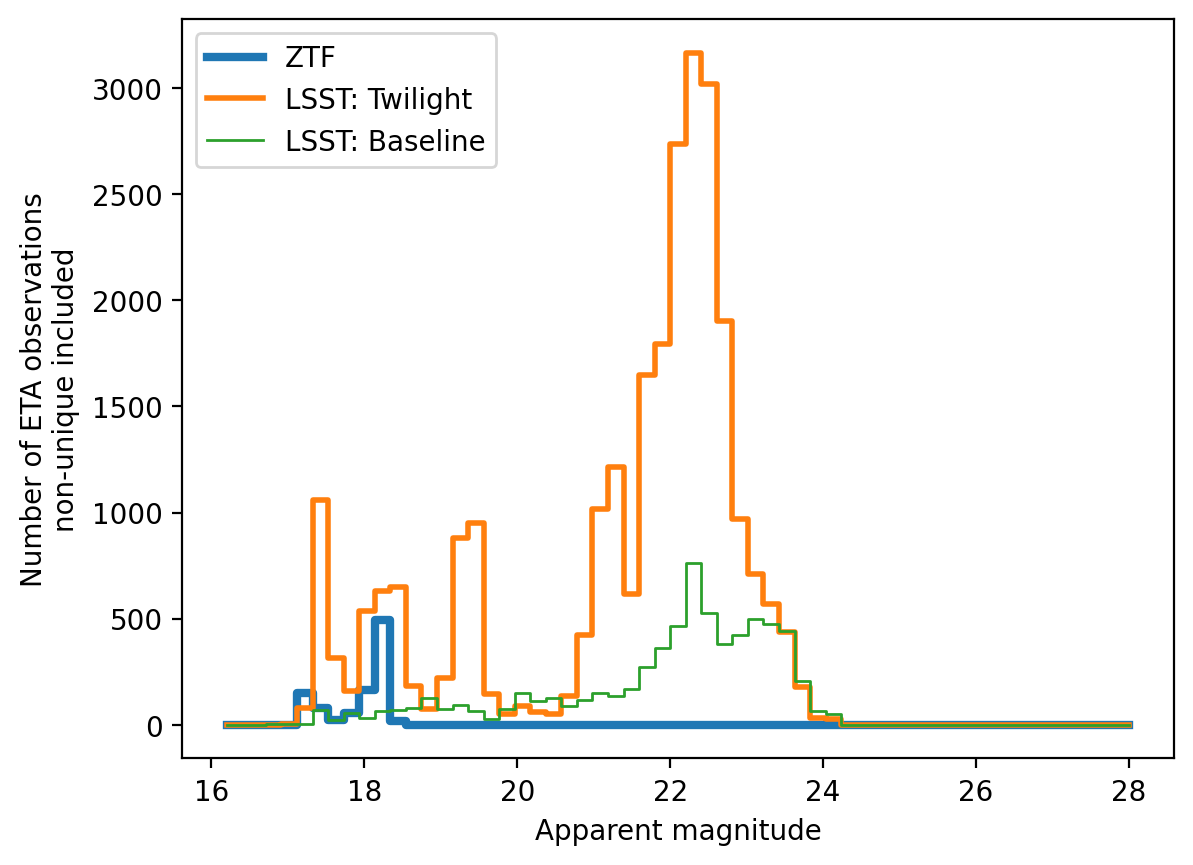}
    \caption{Total \change{number of ETA} detections\change{, as a function of} apparent magnitude, per survey.}
    \label{fig:apparent_magnitude}
\end{figure}

\indent The apparent magnitudes \move{(irrespective of filter)} of the ETAs detected by LSST baseline, \change{LSST} twilight\change{,} and ZTF \change{surveys} are provided in \Cref{fig:apparent_magnitude}. ZTF cannot detect any ETAs beyond an apparent magnitude \change{($m_{app}$)} of 21, while most LSST observations are at apparent magnitudes between $H=$ 21 and $H=$ 23. Detections of ETAs quickly fall off after an apparent magnitude of $H=$ 23 but are still possible by LSST out to $H=$ 24.3. The LSST twilight survey is unsurprisingly much more successful at repeatedly detecting ETAs at $m_{app}>21$. The LSST baseline and twilight are similar in detections of unique ETAs, but twilight produces more than an order of magnitude more repeat detections.

\begin{figure}
    \centering
    \includegraphics[width=0.475\textwidth]{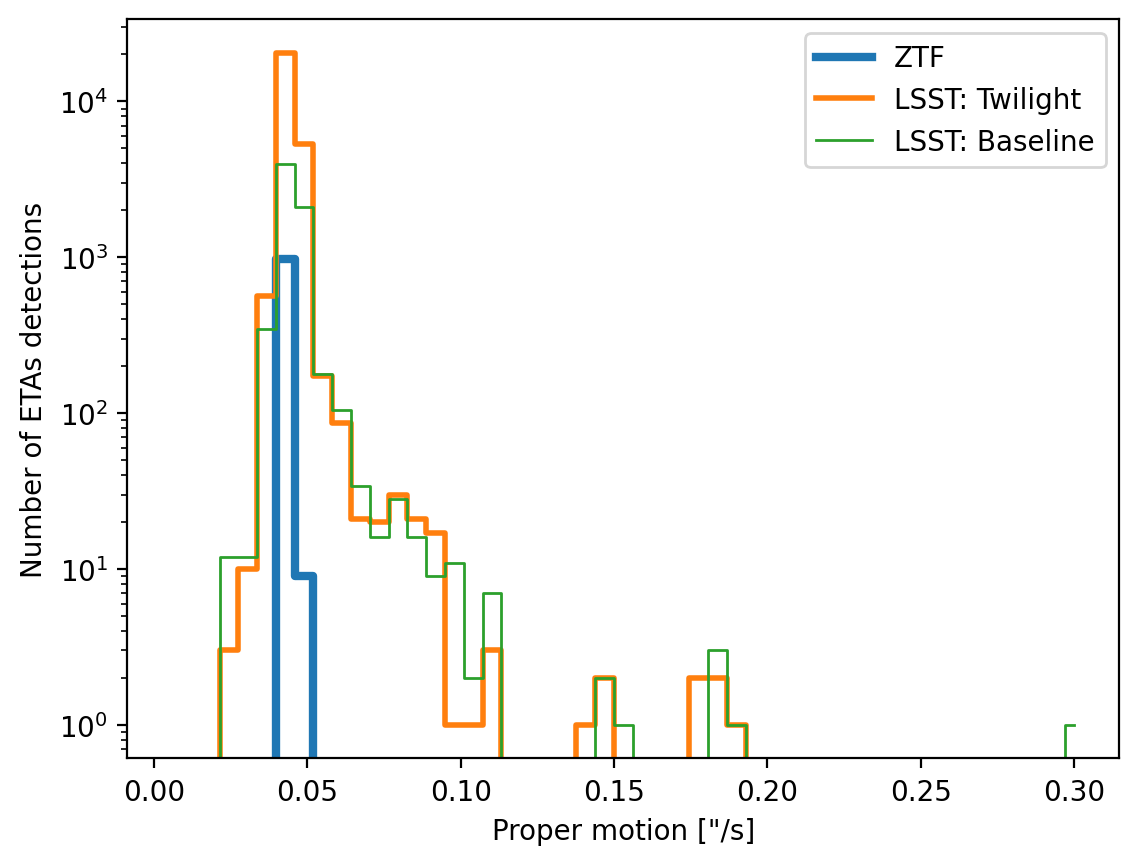}
    \caption{Total \move{ETA} detections\change{, as a function of ETA} proper motion\change{, per} survey.}
    \label{fig:propermotions}
\end{figure}

\Cref{fig:propermotions} shows \change{that} the proper motions of detected ETAs peak\change{s} at 0.05 \change{arcseconds per second,} with an extremely sharp fall-off tail reaching 0.2 \change{arcseconds per second}. The LSST baseline made the fastest ETA detection\change{, at} 0.3 \change{arcseconds per second}. \change{However}, the \change{LSST} twilight survey's overall proper motion distribution is very similar to that of the baseline \change{survey}. The on-sky speed is typical of ETAs at their closest approach to Earth. ZTF, with a larger field\change{-}of\change{-}view, still failed to capture any ETAs moving faster than 0.10 \change{arcseconds per second}. This speaks more to the apparent magnitude limit of ZTF than LSST; if ZTF could make more detections, a similar\change{ly} small fraction would be expected \change{up} to proper motions of 0.3 \change{arcseconds per second}. 

\begin{figure}
    \centering
    \includegraphics[width=0.475\textwidth]{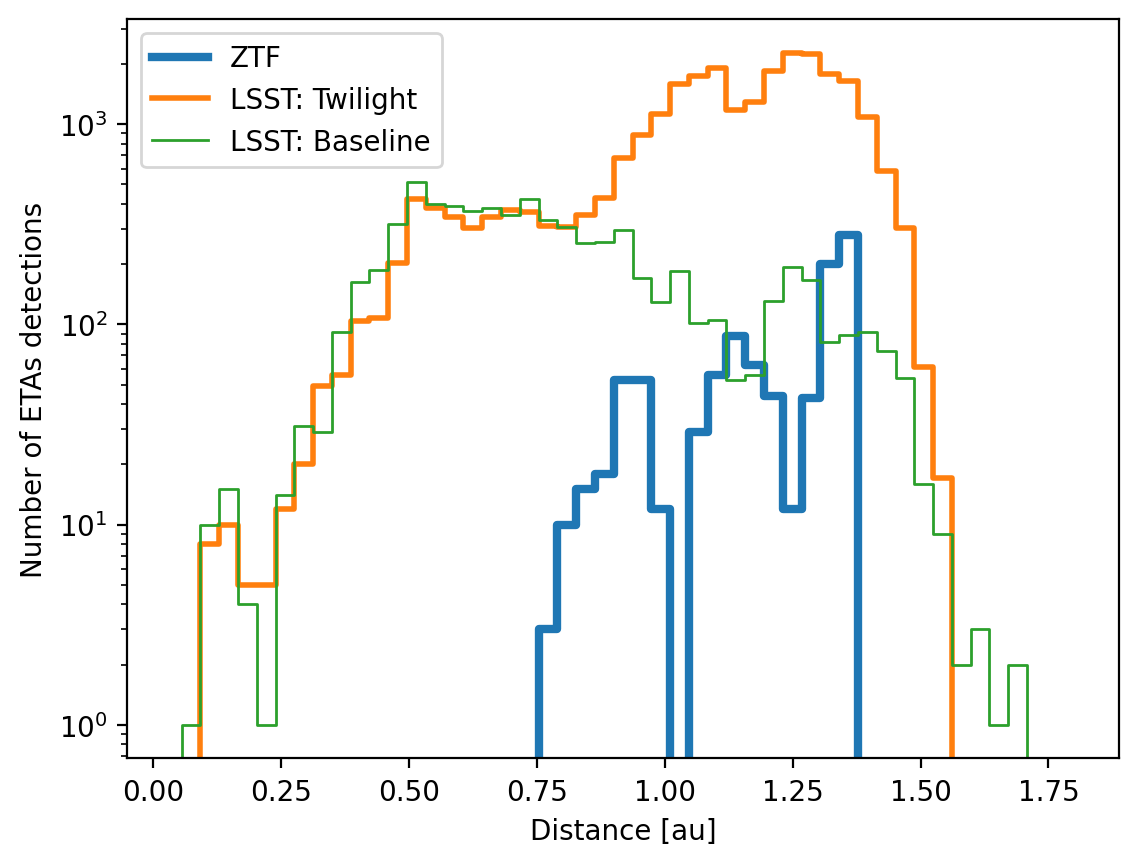}
    \caption{All detected ETAs\change{, as a function of} distance from the Earth \change{(}in units of \change{AU), per survey}.}
    \label{fig:distance}
\end{figure}

\indent \Cref{fig:distance} shows the distance an ETA is from Earth at the time of detection. ZTF can make ETA detections out to 1.7 AU, the practical limit due to the angle of \change{S}olar elongation becoming too small. As seen in the X-Y map of ETA trajectories in \Cref{fig:fullstackxy}, ETAs travel out to L3 \change{(}behind the Sun as viewed from Earth\change{)}, though detections cannot be made from Earth beyond a distance of 1.75 \change{AU,} as the line of sight is too near the Sun and neither LSST nor ZTF point at solar elongations less than 36\degree{}. ZTF detects $\sim$5-10x fewer ETAs\change{,} at all distances\change{,} compared to LSST's baseline and twilight survey\change{s}. The LSST baseline is nearly 1:1 with the twilight survey for ETA detections \change{at} 0.75 \change{AU} and closer. However, a significant deviation in detections occurs \change{beyond} 0.75 \change{AU}, where the LSST twilight survey becomes much more successful at detecting ETAs. The ZTF survey probes a few degrees closer to the Sun than the LSST twilight survey\change{,} allowing a few more detect\change{ions} at distances out to 1.75 \change{AU}.

\begin{figure*}
    \centering
    \includegraphics[width=0.9\textwidth]{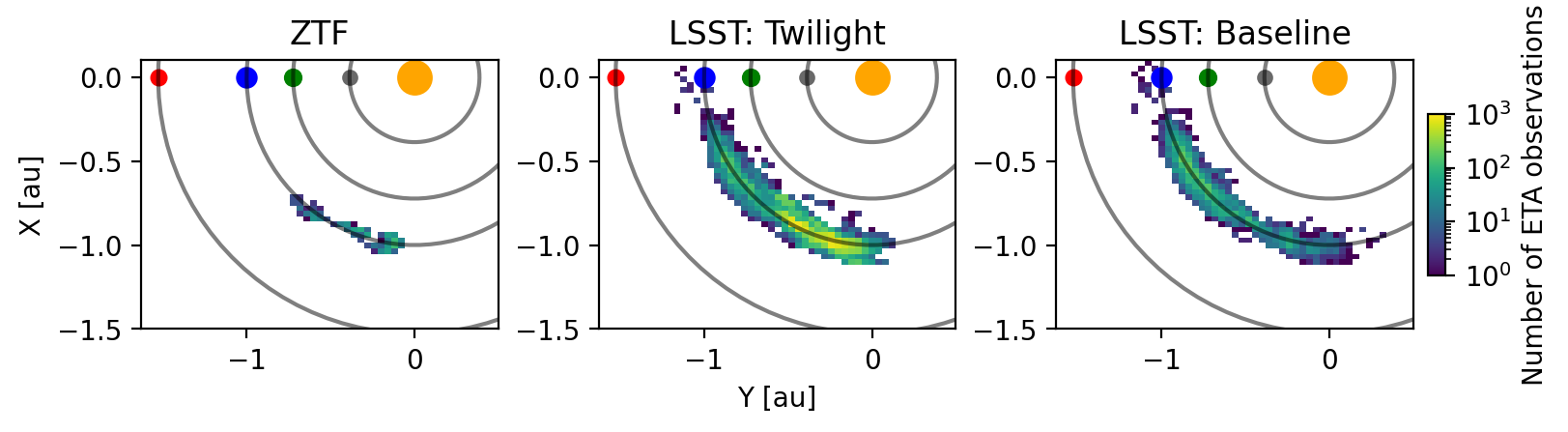}
    \includegraphics[width=0.9\textwidth]{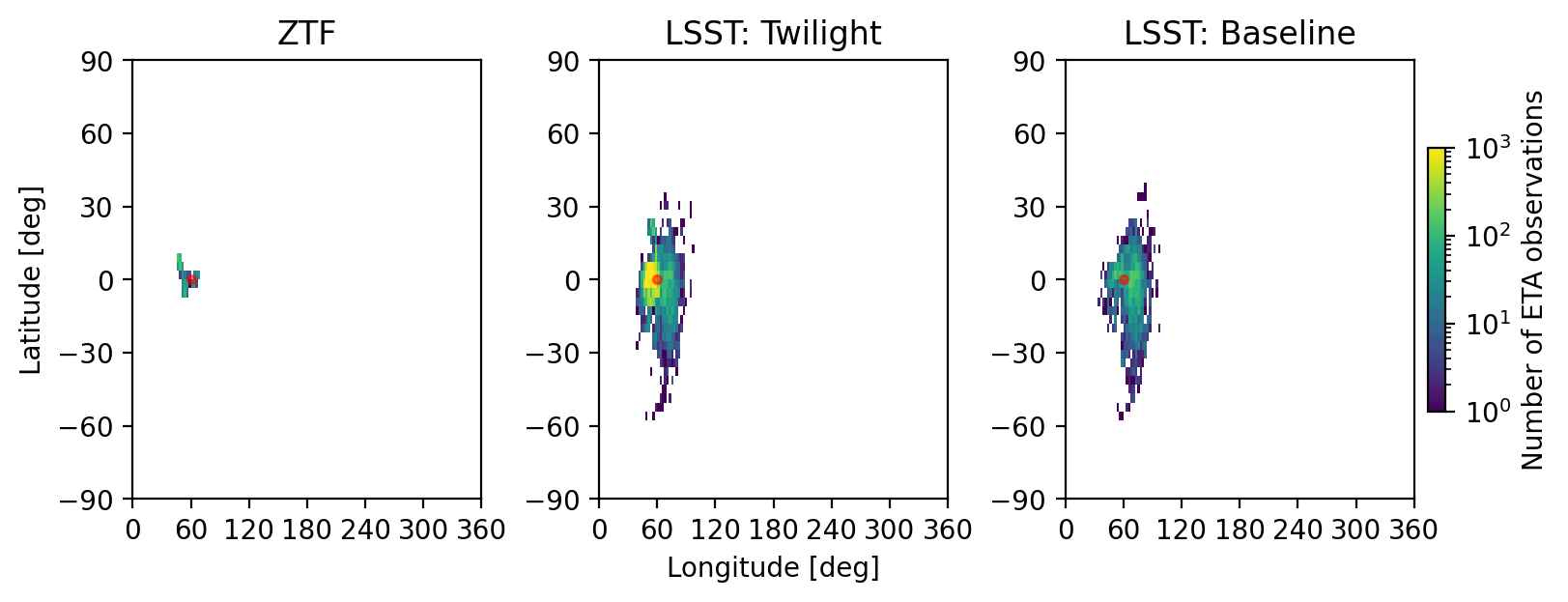}
    \includegraphics[width=0.9\textwidth]{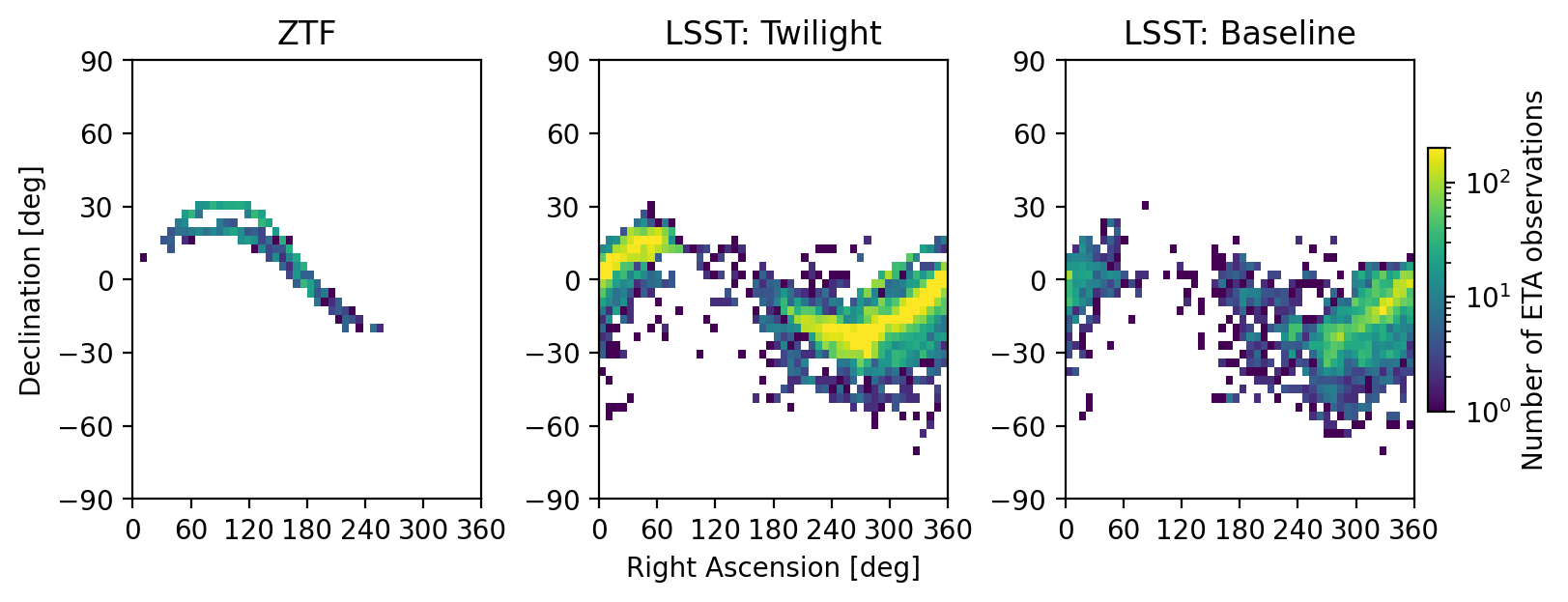}

    \caption{2\change{D} histograms of detected ETA distributions. \change{Values are binned into a} 50x50 \change{grid}. The three columns from left to right are for \change{} ZTF, LSST twilight, and LSST baseline surveys. The three rows are: 
    1) X vs. Y; top-down view of the \change{S}olar \change{S}ystem with the Sun at (0,0) and from the left \change{to right in the grey circles are} the approximate orbits of Mars, Earth, Venus, and Mercury. The frame is such that the position of Earth is fixed to the point (-1,0). The markers for the other planets are not their true position\change{,} but denote the orbits.
    2) Latitude vs. longitude; this is an on-sky coordinate system viewed from Earth, the same as used in \Cref{fig:fullstacklatlon}. The Sun is fixed at (0\degree,0\degree), and L4 is at (0\degree,60\degree), denoted by a red dot.
    3) R.A. vs. Dec.
    }
    \label{fig:maps}
\end{figure*}

\indent \Cref{fig:maps} shows three different views (one in each row) of the ETA detection distributions for the three surveys (one in each column). The top row shows the distribution of ETA detections viewed from the pole of the Solar System. The middle row shows the distribution as viewed from Earth, where (0\change{\degree{}},0\change{\degree{}}) is the position of the Sun and (0\change{\degree{}},60\change{\degree{}}) is the L4 point. The bottom row shows the R.A. and Dec. of the ETA detections. All three surveys can detect ETAs within similar X-Y bounds, as seen in the top row of the figure. The detections made by ZTF all occur closer to the orbit of Earth, with approximately an even number of observations made over all distances from Earth. This is due to ZTF only picking up a few relatively bright ETAs\change{,} but observing them repeatedly over a large libration about L4. Comparing the LSST baseline \change{and} twilight results, a stark difference is seen in the X-Y detection distribution\change{,} as the twilight survey successfully detects many more ETAs at solar elongations of 30-35\change{\degree{}} \change{(}which correspond to distances of 0.75 \change{AU} to 1.7 \change{AU} \change{away} \move{from Earth}. The increased bright region in \change{the} middle \change{column} of the first two rows \change{of} \move{\Cref{fig:maps}} corresponds to the \change{increased number} of twilight detections \change{between 0.75 and 1.50 AU} in \Cref{fig:distance}. The ability \change{for} a survey to make more observations at solar elongations within L4 significantly increases detections of ETAs. The R.A. and Dec. distribution follows the ecliptic but also re-highlights that ZTF observes ETAs between June and December, and LSST, because of its southern latitude, observes the L4 ETAs from January to late June.

\begin{figure*}[htb!]
    \centering
    \includegraphics[width=0.475\textwidth]{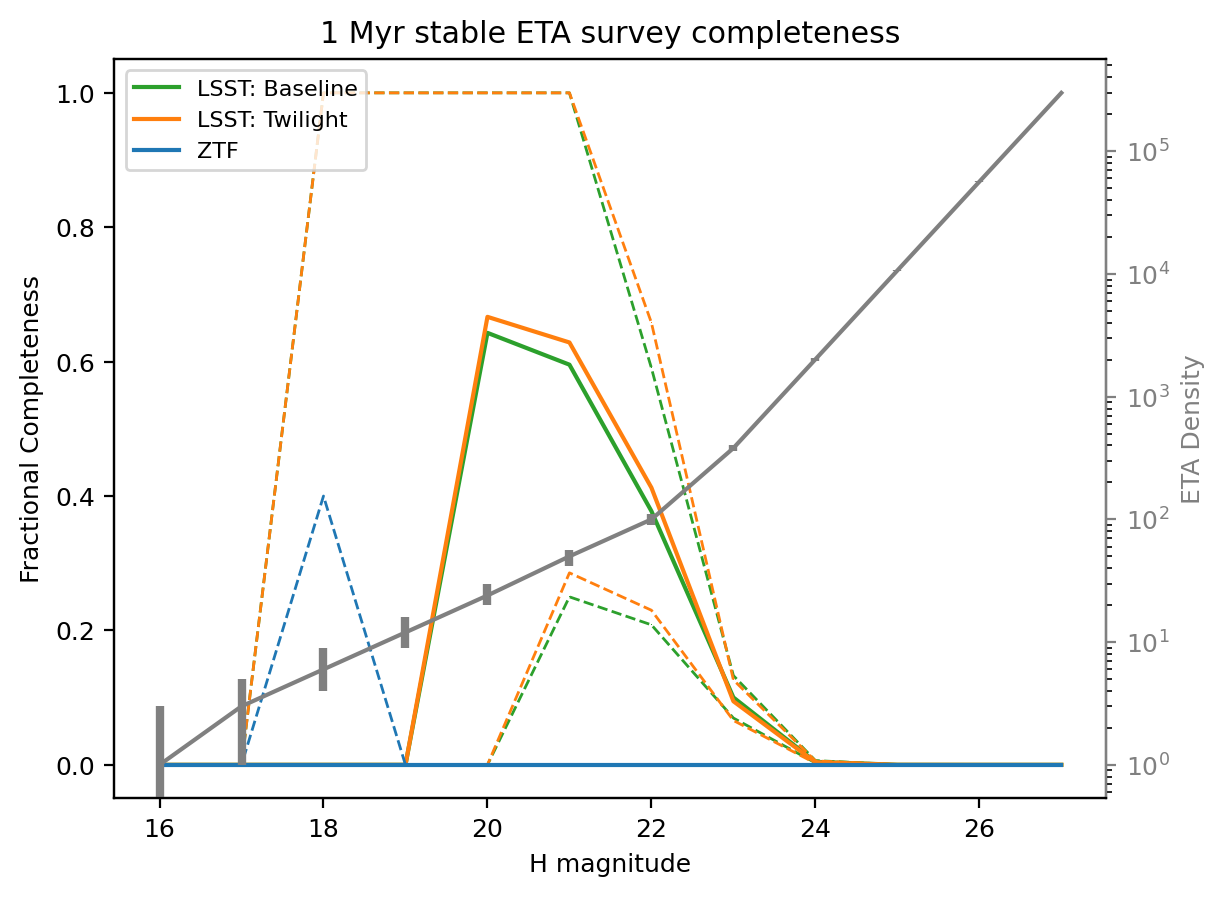}
    \includegraphics[width=0.475\textwidth]{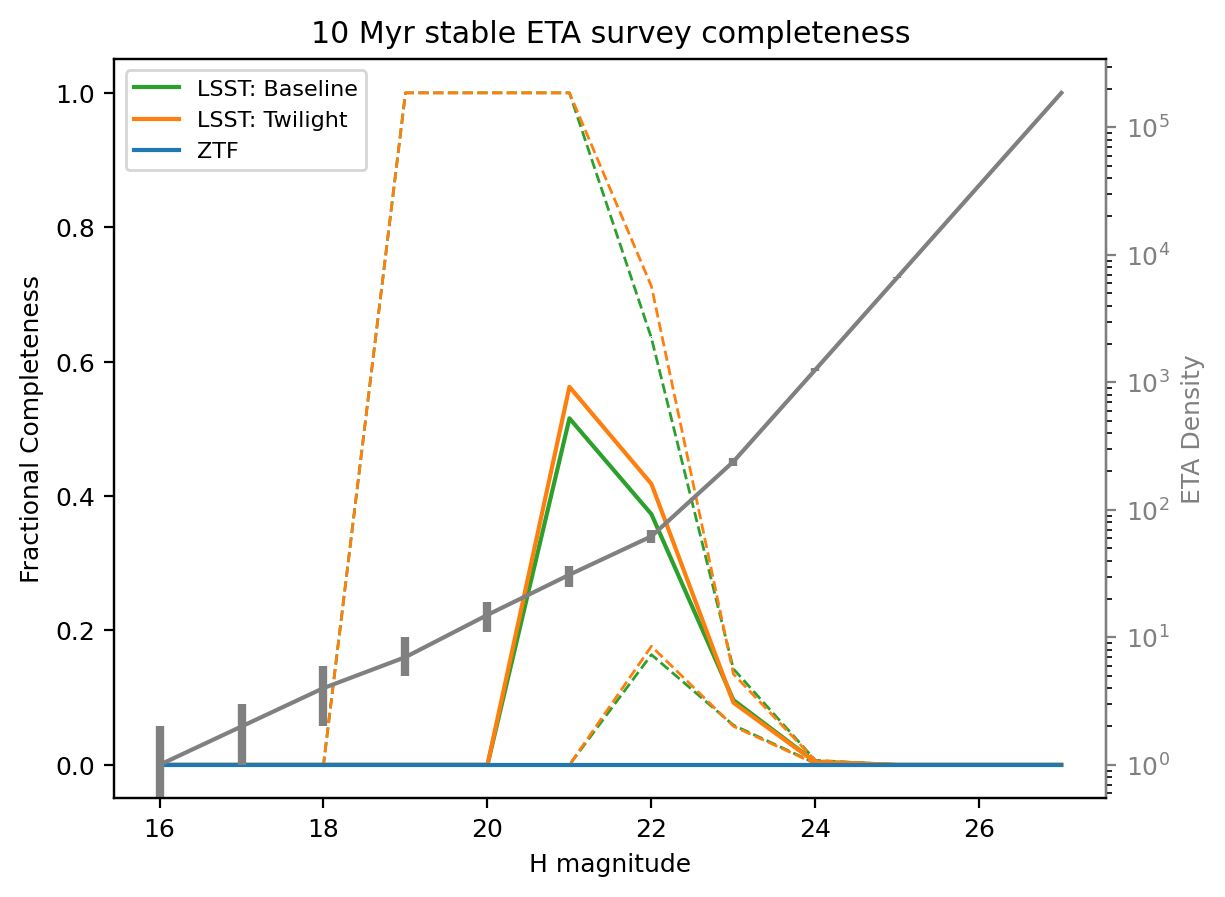}
    \includegraphics[width=0.475\textwidth]{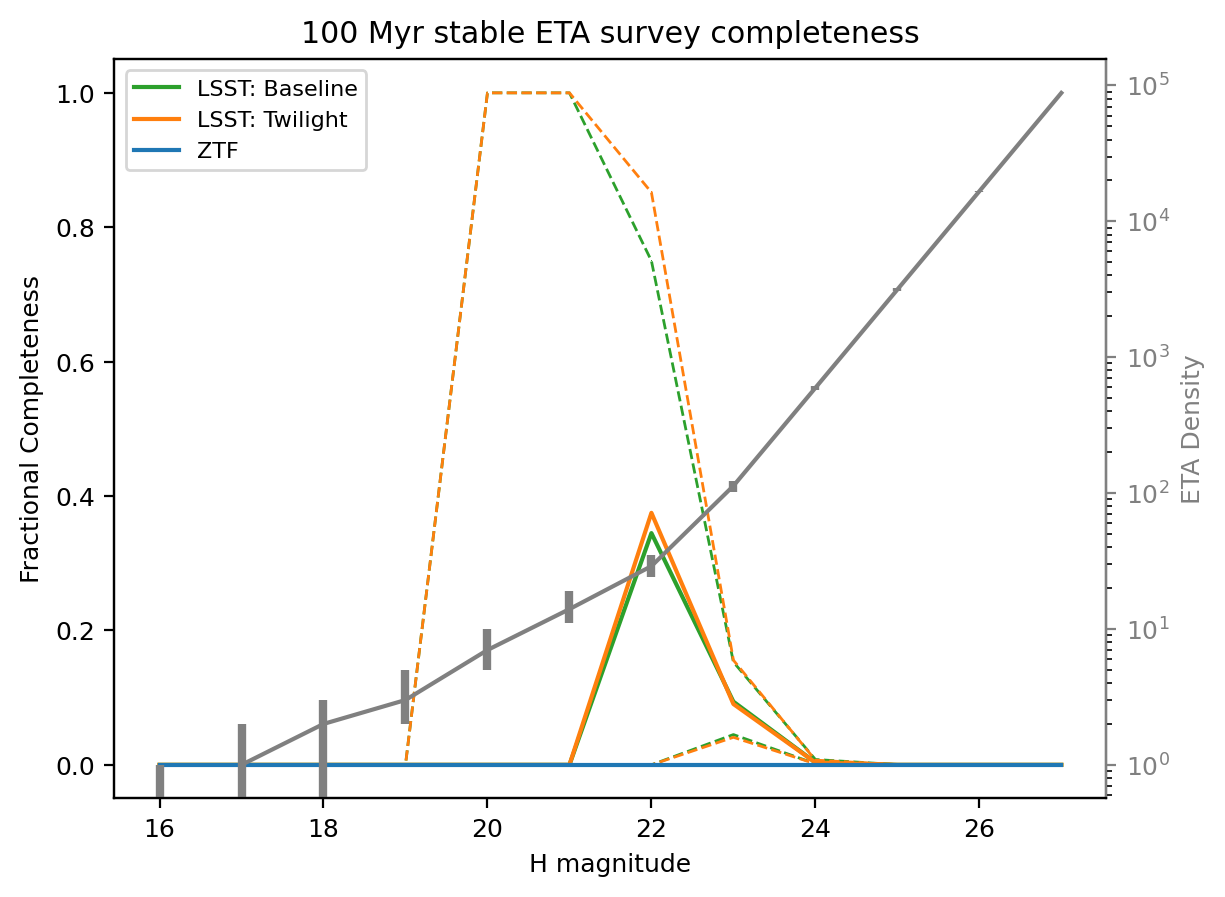}
    \includegraphics[width=0.475\textwidth]{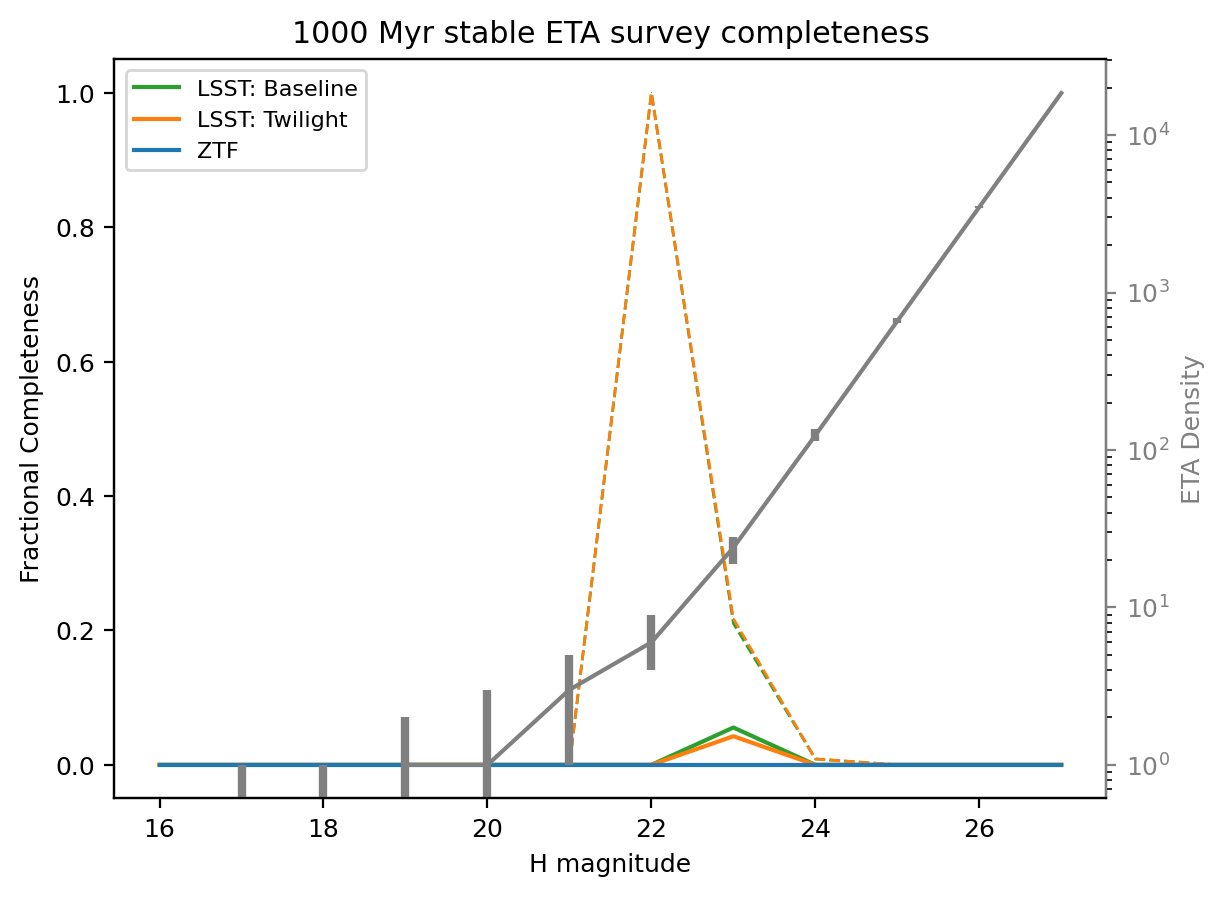}
    
    \caption{The fraction of simulated ETAs detected in each survey. In green is the LSST v2.2 baseline, orange is the LSST twilight, and blue is ZTF. The gray line represents the median ETA absolute magnitude distribution, with a 68\% CI shown as error bars. The \move{68\% CI} fractional survey completeness is shown \change{by the} dotted lines. The fainter end of these distributions will weaken due to streaking.}
    \label{fig:fraction_detected}
\end{figure*}

\subsection{\change{D}etectable \change{F}raction of the ETA \change{P}opulation by LSST and ZTF}

\indent The fraction of ETAs detected \change{per} \move{survey} was determined for 1000 bootstrap samples of the \change{respective} ETA populations. In \Cref{fig:fraction_detected}, the survey completeness for ETA detections as a function of $H$ is shown. The gray line represents the median simulated ETA $H$ distribution, with the 68\% confidence interval (CI) shown as error bars. The median survey completeness is shown as solid color lines. The 68\% confidence intervals are shown as dotted lines. Low number statistics at the bright end of the $H$ distribution tend to cause an all-or-nothing detection of ETAs \change{( resulting in} wide error bars \change{between 0 and 1)} for the survey completeness\footnote{\change{Larger asteroids correspond to lower magnitudes, but larger asteroids are more rare, meaning you aren't always likely to get a larger asteroid in the resample. If you do get a large asteroid you will almost certainly detect it, but if you don't then you are unlikely to get that detection. This leads to this `all-or-nothing' error profile, but reflects the rarity of large asteroids, rather than defects in the observational surveys.}}. The \change{upper} left panel \change{illustrates} that the median number of detected ETAs for each survey is zero \change{for} $H<18$\change{,} due to the fact that brighter ETAs are very rare. \change{However, this} does not suggest that LSST and ZTF would fail to detect such bright ETAs\change{,} but rather that the occurrence of a bright ETA within the survey footprint is rare in our bootstrap sampling procedure. As the simulation time progresses, the low number statistics become more prominent for smaller and smaller (originally more numerous) ETAs. In the bottom right panel, no ETAs with $H<19$ are present in any of the bootstrapped populations. 

\indent The subpanels of \Cref{fig:fraction_detected} show the fractional completeness for a series of different stability subgroups of ETAs that were selected by their lifetime in the MEGASIM. ETAs were removed from the MEGASIM if they crossed the plane perpendicular to the ecliptic that intersects the Earth and L3. In the MEGASIM, only gravitational forces were used to compute the motion of the ETAs. Additional forces, such as thermal effects like the Yarkovsky effect, would, in reality, further reduce the number of ETA trajectories that remain on long timescales \change{(}especially for smaller asteroids\change{)}. For orbits that persist on Myr timescales, additional forces are negligible. If an orbit persists on the order of 1 Gyr, ETAs with smaller diameters ($\sim100$ meters) will be pushed out of \change{their} tadpole\change{-shaped} stability regime\change{s} \citet{Zhuo2019A&A...622A..97Z}. The corresponding absolute magnitude for 140 meter asteroids is approximately $H=20.1$ and $H=22.2$\change{,} for \move{C-type and} S-type ETAs\change{, respectively}. In the \change{lower} right panel \change{of \Cref{fig:fraction_detected}}, the only detections that remain are fainter than these limits.

\subsection{Estimating the \change{U}pper \change{L}imit of the ETA \change{P}opulation}
\begin{figure*}
    \centering
    \includegraphics[width=0.475\textwidth]{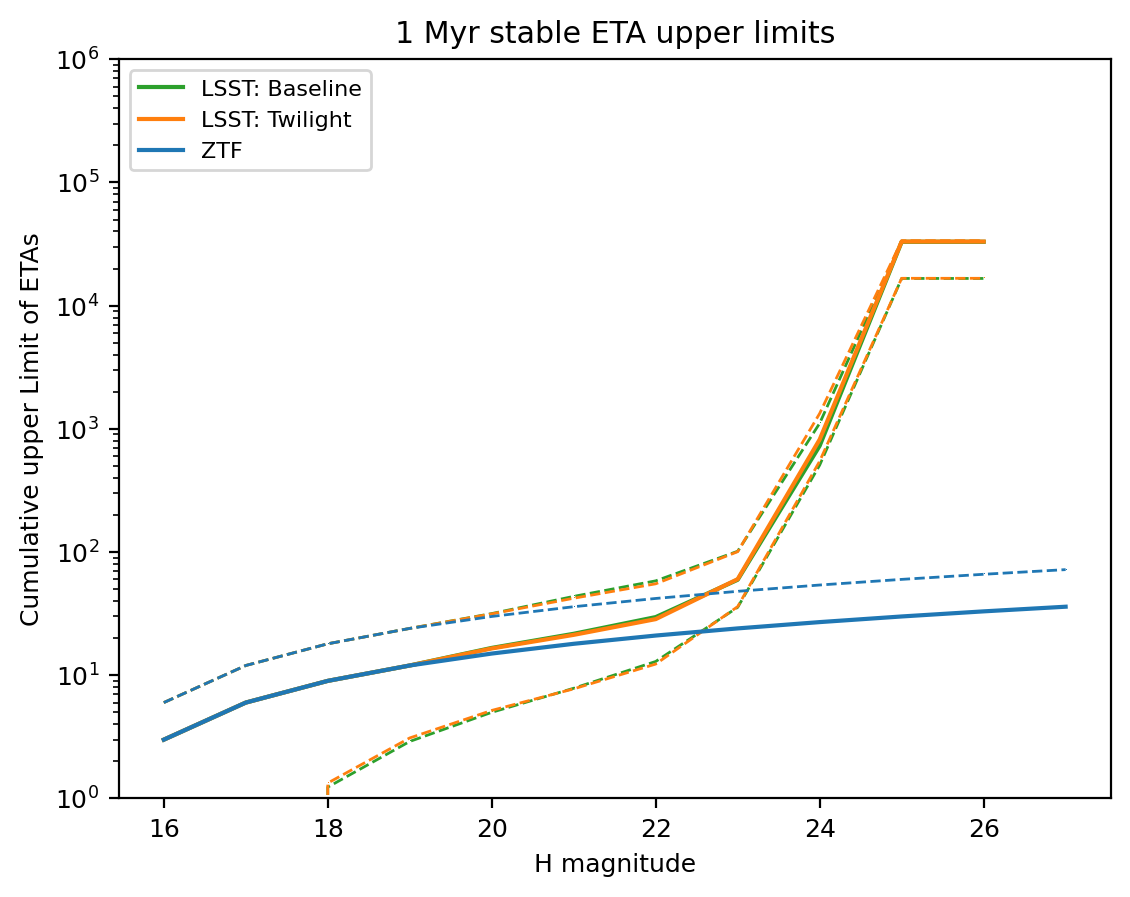}
    \includegraphics[width=0.475\textwidth]{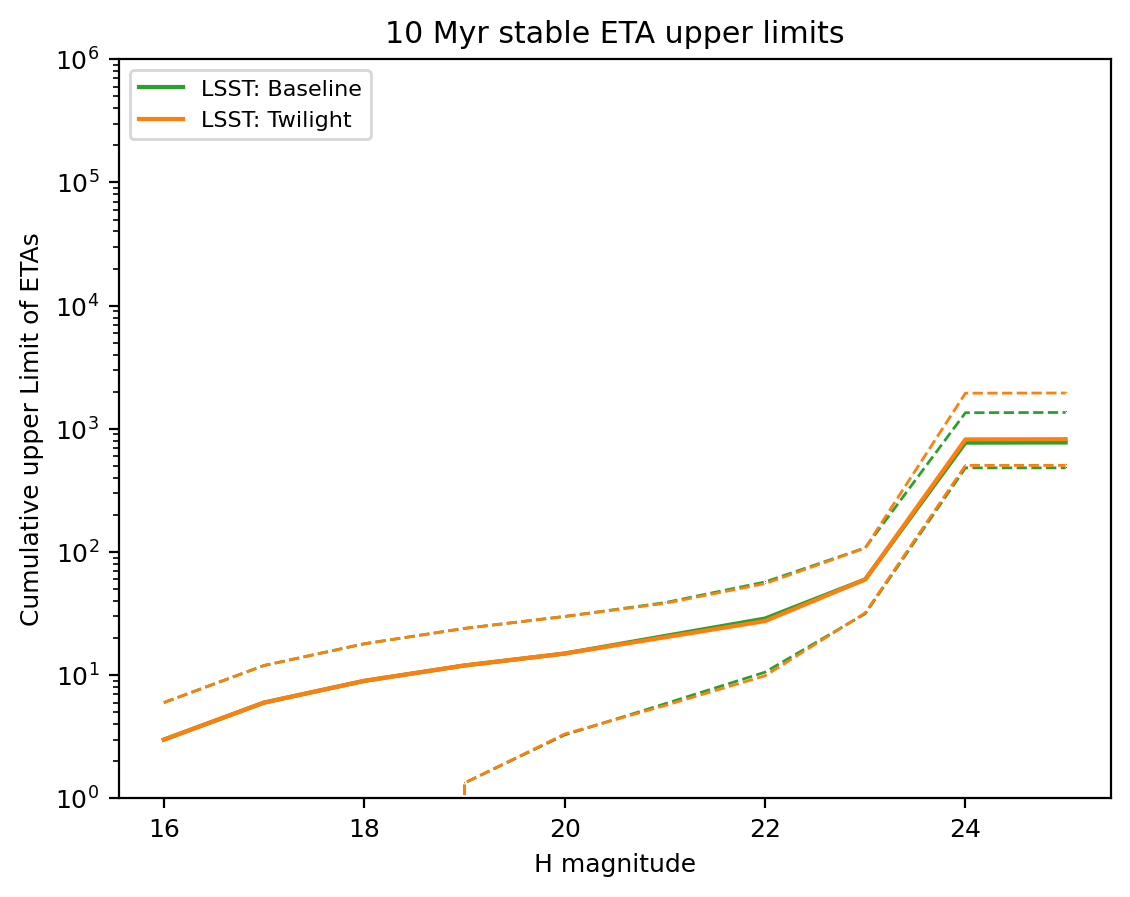}
    \includegraphics[width=0.475\textwidth]{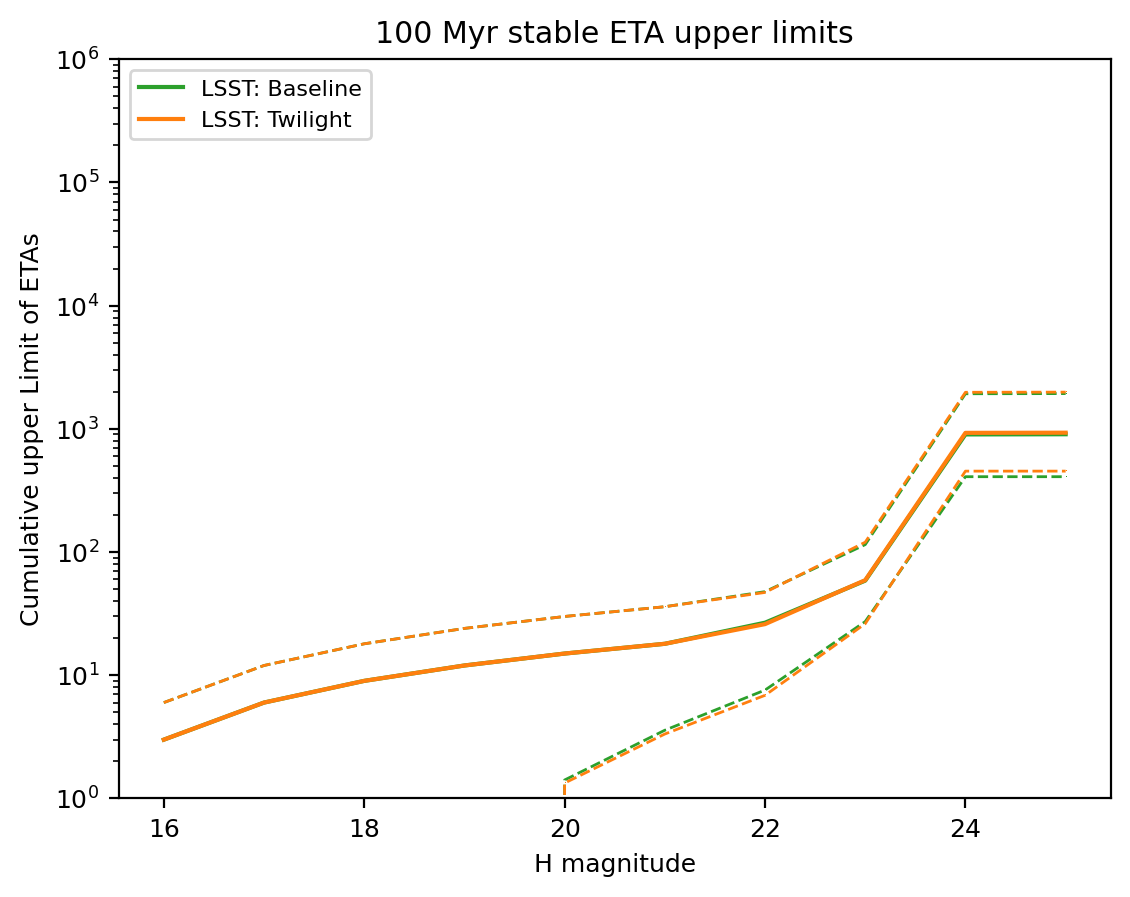}
    \includegraphics[width=0.475\textwidth]{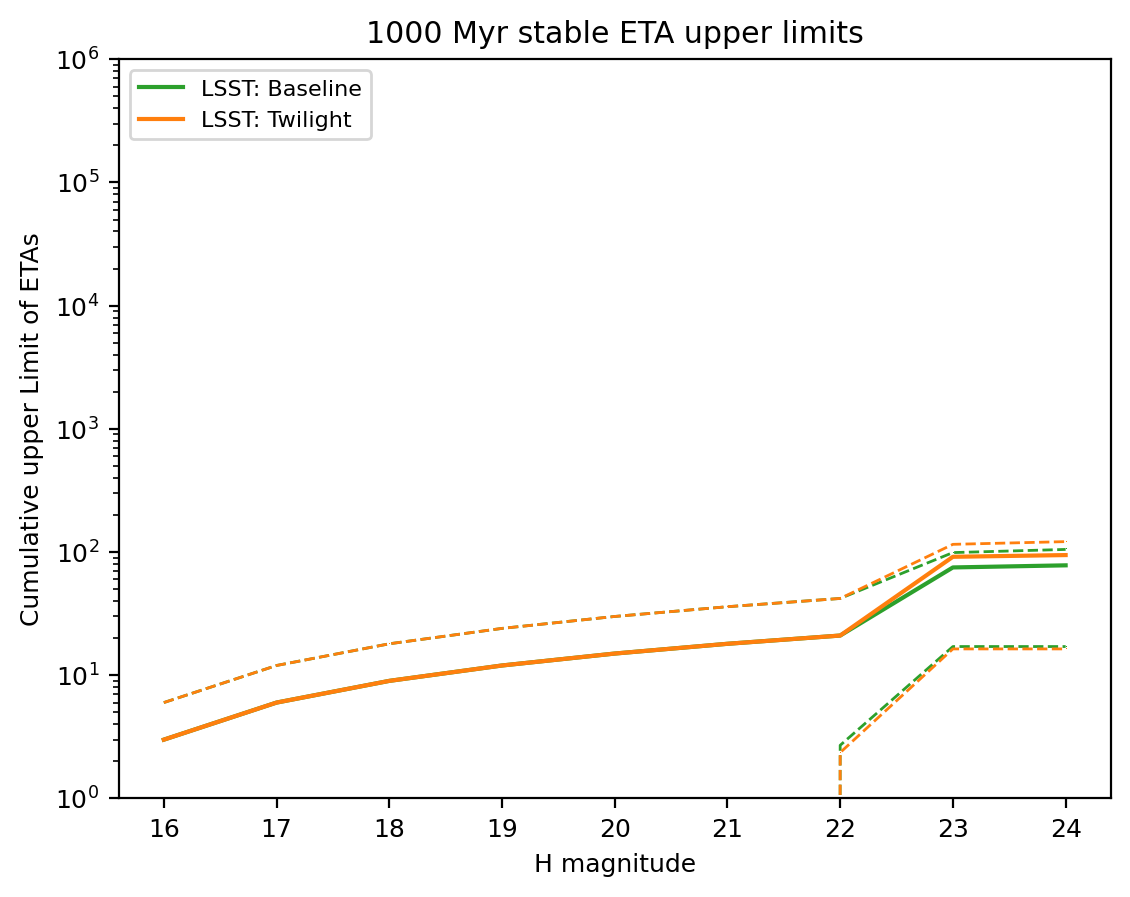}
    
    \caption{Cumulative upper limit on ETA populations for the three surveys. The dotted lines show the 68\% CI.}
    \label{fig:uh}
\end{figure*}

\indent \Cref{fig:uh} is the cumulative upper limit for the ETA population determined via the fractional completeness curve. Following the method of \citet{2020MNRAS.492.6105M}, we determine the upper limit population allowable assuming a null detection of ETAs in each survey. \move{\change{We assume a null detection for ZTF, meaning we do not believe ZTF has detected any ETAs. Borrowed from the aforementioned method, we also consider a null detection f}or LSST, \change{where the simulated} detection is \change{considered} within 3$\sigma$ \change{(standard deviations)} of the simulation results (which happens out to H=19 for the twilight survey\change{,} and H=20 for the baseline survey).} We calculate the upper limit using \Cref{eq:uh}, where $U(H)$ is the upper limit at a given absolute magnitude, $f(H)$ is the fraction of ETAs detected at a given absolute magnitude, and the 3 in the numerator implies the upper limit is within 3$\sigma$ of the null detection. 
\begin{equation}
\label{eq:uh}
    U(H) = \frac{3}{f(H)}
\end{equation}
\Cref{fig:fraction_detected} shows $U\left(H\right)$ for the same four stability regimes shown previously. In \Cref{fig:fraction_detected}, there are portions of the completeness curves that are zero, which results in infinite error bars. To handle this, undefined lower or upper error bars \change{resulting from} divi\change{ding} by zero are replaced by -3 or +3, respectively.

\subsection{The Yarkovsky Effect and ETA Detections}


\indent In the MEGASIM, only gravitational forces were used to propagate the ETA orbits. Here we estimate the number of gravitationally stable ETAs that may be driven unstable by non-gravitational effects. The dominant additional force to consider is the Yarkovsky Effect \citep[e.g.,][]{2003Sci...302.1739C}, which causes asteroids to rotate and increase their semi-major axis due to  Solar radiation pressure. Simulations of ETA orbits with the Yarkovsky Effect included were done by \citet{Zhuo2019A&A...622A..97Z}. The Yarkovsky Effect's strength depends on the shape, albedo distribution, surface conductivity, spin, and density of a given asteroid. The additional force from the Yarkovsky Effect was found to remove ETAs within 1 Gyr\change{,} for ETAs with diameters smaller than 130 m\change{eter} \change{or} 90 m\change{eter,} for prograde and retrograde rotation, respectively \citep{Zhuo2019A&A...622A..97Z}.

\indent To estimate the changes to our survey completeness curves that the Yarkovsky Effect would impose, a cut-off on the simulated ETA population was made using both the lifetime and diameter of a given ETA. To estimate the removal rate of gravitationally stable ETAs \change{via the Yarkovsky Effect} in the MEGASIM, we use \change{an ETA diameter cutoff ($D_\text{unstable}$) provided by} \Cref{eq:drift}. 

\begin{equation}\label{eq:drift}
    D_\text{unstable} = \frac{lifetime_{100m}}{1\,\,Gyr}
\end{equation}where \change{$lifetime_{100m}$} is the time a \change{100 meter diameter} ETA remained bound to L4 in the MEGASIM \change{for one} Gyr. \Cref{eq:drift} is a simple approximation for the ETA diameter cutoff, given the results of \change{\citet{Zhuo2019A&A...622A..97Z}, who estimate that ETAs of diameter 100m (on average, given prograde and retrograde spins) will be driven unstable, and removed from the gravitational potential well (thus no longer being Earth Trojan asteroids)} by 1 Gyr. \change{We assume this linear approximation to estimate the number of smaller ETAs affected on shorter time scales}. \change{If an ETA is bigger than this $D_\text{unstable}$ cutoff, it remains stable, while an ETA with a smaller diameter is assumed} \move{to be removed from the population, and therefore, not detectable}. 

\indent In a few percent of the resampled runs\change{,} some ETAs with lifetimes greater than 750 Myr \change{should have been} removed \change{due to the} \move{Yarkovsky Effect}. \change{However, t}he number of ETAs detected in the resamples\change{,} within a confidence interval of 68\%\change{,} were \change{essentially} unaffected by the Yarkovsky effect\change{, as it} \move{is not strong enough to drastically change our estimated survey detections} \change{at these timescales}. \change{This is due to the fact that our detected ETAs are either too large (i.e., t}\move{he diameter of ETAs just large enough to be detected by LSST are also just above the diameter for which the Yarkovsky effect will produce orbital instability}\change{)}, or \change{have not been around long enough for the Yarkovsky Effect to manifest instability (i.e. Gyr timescales).} ZTF is not included because there are no timescales where ZTF detections remain \change{stable} after this procedure. We reiterate that this is due to small number statistics in our bootstrap populations \change{throughout} the detection efficiency analysis above.

\section{Conclusion}\label{sec:conclusions}

\indent The work presented here can be categorized into three areas. 1) The spatial distributions of MEGASIM ETA trajectories, 2) the detectability of MEGASIM ETAs by LSST and ZTF, and 3) an upper limit of an actual ETA population assuming a null detection.

\subsection{ETA \change{S}patial \change{D}istributions}

\indent When viewing L4 ETAs from Earth, the highest on-sky concentration is found 6.5\degree{} closer to the Sun than the L4 Lagrange point. 30\% of ETA trajectories are found within 45-60\degree{} \change{S}olar elongation and another 30\% are found beyond 60\degree{}. 70\% of all ETAs exist within 5\degree{} of the ecliptic plane across all longitudes. \change{Detection} of ETAs from the ground \change{is} challenging inside L4\change{, and d}etections cannot be made at all times of the year due to the Earth's axial tilt \citep{WHITELEY1998154}\change{. O}bservations require longer exposure times due to higher sky brightness at twilight\change{,} and higher airmass than typical observations at opposition.

\indent The distribution of ETAs as viewed looking down the pole of the ecliptic plane shows a toroidal volume that tapers near the Earth and L3 Lagrange point. ETAs can librate in as far as the orbit of Venus\change{,} and out \change{as far as} 1.3 \change{AU,} especially if they are only stable for short periods of time. ETAs \change{that are} \move{stable longer} mostly stay within 0.98 - 1.02 \change{AU} of the Sun\change{, and i}n the ecliptic plane, the highest density of ETAs oscillates over a considerable distance. High densities occur along a ridge \change{spanning} 45\degree{} leading Earth's orbit to 90\degree{}\change{,} with the peak within a few degrees of the L4 point \change{(}if averaged over time\change{)}. \change{ETAs have long libration periods (decades to millennia), so the time it takes to oscillate from behind the Sun to closer toward the Earth could fall completely outside of the survey observation window, meaning that modern surveys for observing the ETA population may not be feasible.}. 

\indent As the lifetime of ETAs increases, the region that surviving ETAs \change{traverse} tightens onto the Earth's orbit. ETAs \change{that are stable over long periods of time} do not approach Earth or L3 as closely\change{,} and remain in a tighter bunch around L4. This is one reason for fewer detections \change{of} longer-term stable ETAs in Figure \ref{fig:fraction_detected}, even for ZTF and the LSST twilight survey, which are notionally designed for such low Solar elongation observations. 

\indent The on-sky location where ETAs are brightest does not correlate to the most likely position on the sky. The brightest on-sky areas are found at \change{S}olar elongations outside of L4\change{,} at angles of 60\degree{} to 75\degree{}. The brightest location is on the ecliptic, but elevated brightness regions spread beyond L4 to latitudes of $\pm20$. This is apparent in \Cref{fig:differencelatlon} as a flaring yellow regime to the right of L4, and is due to the improved observing geometry. 

\subsection{Ability of LSST and ZTF to \change{D}etect ETAs}

\indent Due to the latitude that \change{the telescopes for} ZTF and LSST are located, the time of year \change{in which} each can detect L4 ETAs differs (see \Cref{fig:detectionspermonth}). ZTF finds most ETAs from June to December\change{,} and LSST from January to June. In our simulations, ZTF was unsuccessful at detecting the vast majority of the ETA population\change{,} due to limited photometric sensitivity ($H<19$ for most detections). Of the ZTF detections, only 4 unique ETAs were observed. Though, \change{due to} the field\change{-}of\change{-}view of the ZTF telescope\change{,} repeat detections are on the order of thousands, which indicates that any detected ETA by ZTF would be very well characterized and tracked.

\indent Both the LSST baseline and twilight survey recovered \new{the same number of unique ETAs\change{,} \move{188}.} \move{However,} \change{t}he twilight survey was able to make several thousand repeat detections, with \change{4} of the asteroids observed over 10,000 times. These results show that a twilight survey will not increase the likelihood of individual ETA detections drastically compared to the baseline survey. Though, a twilight survey would provide us with several orders of magnitude more observations of select ETAs. Observing asteroids on the order of thousands to tens of thousands of times is unprecedented and would be useful for constraining characteriz\change{ation of} individual asteroids.

\subsection{The \change{P}ossible ETA \change{P}opulation}
\indent ZTF did not have a statistically significant number of detections to provide meaningful constraints on the upper limit of the ETA population\change{, for any size range. That is, at magnitude $H<$ 18, the probability that an ETA exists become so low that ZTF’s observing cadence will not detect the ETA frequently enough to count as an observed ETA.} However, both the LSST baseline and twilight surveys provided enough detections to estimate an upper limit for ETAs\change{,} of Myr up to Gyr stability times. Due to the Yarkovsky Effect, C-type and S-type ETAs of absolute magnitudes greater than $H=22$ \change{and} $H=24$\change{, respectively,} will be driven unstable within 1 Gyr. The rarity of ETAs greater than 100 m\change{eters} in diameter\change{,} paired with the fact that smaller asteroids will be driven unstable, our results indicate that a null detection by LSST will mean that there are no remaining Gyr-stable or primordial ETA populations. On the other hand, the existing ETA discoveries are indicative of a small and transient population. A null detection in LSST will restrict that population to tens of objects larger than 100 meters.

\section*{Acknowledgements}
\indent We would like to thank Bryce Bolin for compiling the ZTF pointings information and helping us decipher the catalog for our purposes here. We thank Michael Schneider for valuable discussions \new{and Alexx Perloff for his help on validating code}.

\indent Computing support for this work came from the Lawrence Livermore National Laboratory Institutional Computing Grand Challenge program. This work was performed under the auspices of the U.S. Department of Energy by Lawrence Livermore National Laboratory under Contract DE-AC52-07NA27344 and was supported by the LLNL-LDRD Program under Projects 20-ERD-025, 23-ERD-044, and 22-ERD-054.

\indent This material is based upon work supported in part by the National Science Foundation through Cooperative Agreement AST-1258333 and Cooperative Support Agreement AST-1202910\change{,} managed by the Association of Universities for Research in Astronomy (AURA), and the Department of Energy under Contract No. DE-AC02-76SF00515 with the SLAC National Accelerator Laboratory managed by Stanford University. Additional Rubin Observatory funding comes from private donations, grants to universities, and in-kind support from LSSTC Institutional Members.

\indent This work was based on observations obtained with the Samuel Oschin Telescope 48-inch and the 60-inch Telescope at the Palomar Observatory as part of the Zwicky Transient Facility project. ZTF-II is supported by the National Science Foundation under Grant No. AST-2034437 and a collaboration including Caltech, IPAC, the Weizmann Institute for Science, the Oskar Klein Center at Stockholm University, the University of Maryland/Joint Space-Sciences Institute, Deutsches Elektronen-Synchrotron and Humboldt University, Lawrence Livermore National Laboratory, TANGO/NTHU, Taiwan, the University of Wisconsin at Milwaukee, Trinity College, Dublin, and IN2P3, France. Operations are conducted by COO, IPAC, and the University of Washington.

\bibliography{main}{}
\bibliographystyle{aasjournal}


\label{lastpage}
\end{document}